\documentclass[aps,prd,10pt,notitlepage,superscriptaddress,nofootinbib,numbers,floatfix]{revtex4-2}

\usepackage[utf8]{inputenc}
\usepackage[T1]{fontenc}
\usepackage{lmodern}
\usepackage{microtype}
\usepackage{amsmath,amssymb,amsfonts,mathrsfs}
\usepackage{bm}
\usepackage{booktabs}
\usepackage{graphicx}
\usepackage[dvipsnames]{xcolor}
\usepackage{hyperref}
\hypersetup{colorlinks=true,citecolor=Purple,linkcolor=Purple,urlcolor=Purple,breaklinks=true}
\usepackage[capitalize]{cleveref}
\usepackage{tikz}
\usepackage{placeins}
\usepackage{flafter}
\usepackage{float}
\AtBeginDocument{\raggedbottom}


\definecolor{lime}{HTML}{A6CE39}
\DeclareRobustCommand{\orcidicon}{%
	\begin{tikzpicture}
		\draw[lime, fill=lime] (0,0)
		circle [radius=0.16]
		node[white] {{\fontfamily{qag}\selectfont \tiny ID}};
		\draw[white, fill=white] (-0.0625,0.095)
		circle [radius=0.007];
	\end{tikzpicture}
	\hspace{-2mm}
}

\foreach \x in {A, ..., Z}{%
	\expandafter\xdef\csname orcid\x\endcsname{\noexpand\href{https://orcid.org/\csname orcidauthor\x\endcsname}{\noexpand\orcidicon}}
}

\newcommand\orcidZijun{{\href{https://orcid.org/0009-0001-1322-6303}{\orcidicon}}}
\newcommand\orcidJonathan{{\href{https://orcid.org/0000-0001-9291-0893}{\orcidicon}}}
\newcommand\orcidEdson{{\href{https://orcid.org/0000-0001-9929-5977}{\orcidicon}}}
\newcommand\orcidMinyong{{\href{https://orcid.org/0000-0001-5577-575X}{\orcidicon}}}

\newcommand{\dd}{\mathrm{d}}
\newcommand{\Fh}{\,{}_2F_1}
\newcommand{\Mone}{\mathcal{M}_1}
\newcommand{\Mtwo}{\mathcal{M}_2}

\begin{document}

\title{Regularization effects on the observational appearance of asymmetric thin-shell wormholes from a new string-cloud geometry}

\author{Zijun Zhong\orcidZijun\!\!}
\email{zjzhong@mail.bnu.edu.cn}
\affiliation{School of Physics and Astronomy, Beijing Normal University, Beijing 100875, China}

\author{Jonathan A. Rebou\c{c}as\orcidJonathan\!\!}
\email{jalvesreboucas@ifce.edu.br}
\affiliation{Instituto Federal de Educa\c{c}\~ao, Ci\^encia e Tecnologia do Cear\'a (IFCE), Iguatu--CE, Brazil}

\author{Edson Otoniel\orcidEdson\!\!}
\email{cedson.otoniel@ufca.edu.br}
\affiliation{Universidade Federal do Cariri (UFCA), Instituto de Forma\c{c}\~ao de Educadores--IFE, R. Oleg\'ario Emidio de Araujo S/N, Brejo Santo--CE, 63.260-000, Brazil}

\author{Minyong Guo\orcidMinyong\!\!}
\email{minyongguo@bnu.edu.cn}
\affiliation{School of Physics and Astronomy, Beijing Normal University, Beijing 100875, China}
\affiliation{Key Laboratory of Multiscale Spin Physics (Beijing Normal University), Ministry of Education, Beijing 100875, China}
\date{\today}

\begin{abstract}

We investigate how rational regularization of the new Letelier--Alencar string-cloud geometry changes the optical appearance of reflection-asymmetric thin-shell wormholes. We compare junctions with two nonregular exteriors, a regular observer exterior paired with a nonregular one, or two regular exteriors at fixed leading asymptotic masses, throat radius, string strength and regulator fractions. Critical scales, ray classes, orbit numbers and transfer functions are combined with three emission profiles and matched black-hole controls. Face-on and inclined images use the same prescribed emitter motion: circular timelike geodesics outside the innermost stable circular orbit and an inward plunge conserving its specific energy and angular momentum inside. Although the core is excised, regularization changes the retained exterior and throat matching. For the sampled parameters, the observer critical screen impact decreases from $10.392\mu_1$ to $9.642\mu_1$, where $\mu_1$ is the observer-side asymptotic mass scale. Regularizing the second exterior then contracts the reflected interval from $3.890\mu_1$ to $1.248\mu_1$ without changing ordinary observer-side rays. Returned brightness depends on source support and velocity-dependent frequency shifts: the regular--regular image nearly coincides with its black-hole control for the outer-edge emission model face-on, while inner emission reveals returned structures. At large sampled inclination that near-degeneracy can be lifted. This separates regularization-dependent propagation from the emission and viewing conditions that determine its image contrast.
\end{abstract}


\maketitle
\tableofcontents
\clearpage
\section{Introduction}\label{sec:introduction}

Horizon-scale imaging has made the propagation of light near compact objects a practical probe of strong gravitational fields. The Event Horizon Telescope images of M87* and Sagittarius A* provide the principal observational motivation, while their interpretation still depends on how spacetime geometry, photon capture, and the emitting plasma are disentangled \cite{EHTM872019,EHTSgrA2022}. The theoretical foundations of this problem extend from escape cones and critical null trajectories to the image of a thin accretion disk and the central brightness depression produced by a compact object \cite{Synge1966,Luminet1979,FalckeMeliaAgol2000}. Analytical studies of shadows and strong lensing consequently remain essential for identifying which image properties follow from the metric and which depend on the source model \cite{PerlickTsupko2022,CunhaHerdeiroRadu2018}.

Photon rings sharpen this separation because near-critical rays can execute multiple angular turns before reaching the observer. Their critical location is controlled by the null geometry, but their width, brightness, and practical prominence depend on transfer-function demagnification and on the distribution of emitting matter \cite{Gralla2019}. The resulting narrow structures have motivated interferometric characterizations and strong-field tests \cite{Johnson2020,GrallaLupsasca2020,GrallaLupsascaMarrone2020}, although the presence of a light ring alone does not determine the nature or stability of an ultracompact object \cite{CunhaBertiHerdeiro2017,CardosoPani2019}. This distinction is particularly important for horizonless geometries, where additional photon paths may exist without producing a bright or source-independent feature.

Traversable wormholes provide a controlled setting in which global connectivity can modify photon propagation while the exterior remains strongly gravitating \cite{MorrisThorne1988}. Thin-shell constructions realize this idea by joining retained spacetime regions across a timelike hypersurface using the junction formalism \cite{Israel1966,Visser1989}. If the two exteriors are not reflection copies, the locally measured photon frequency remains continuous at the shell while the coordinate impact parameters on the two sides generally differ. Circular photon orbits and shell properties may then be linked in ways absent from a single exterior \cite{TsukamotoKokubu2024}, and the opposite region can affect what an observer sees even when both the observer and the emitter remain on the first side.

This reflection asymmetry has already been shown to generate double shadows, shifted critical curves, and additional returned or multiphoton structures in several thin-shell settings \cite{Wang2020,Wielgus2020,Guerrero2021,Tsukamoto2021,TsukamotoDoubleShadows2021}. Thin-disk transfer functions make the mechanism explicit: a ray can cross the throat, turn in the second exterior, return to the observer side, and intersect the disk at image orders not available in the corresponding black-hole exterior \cite{Peng2021}. Thin-disk images of asymmetric thin-shell wormholes have also been studied in Bardeen and renormalization-group-improved geometries, revealing additional photon rings and emission-dependent brightness structures \cite{He:2024yeg,Chen:2024wtw}. Studies incorporating torsion and charge or noncommutative corrections further show how these geometric modifications shift the extra rings and alter their brightness distributions \cite{Wang:2025opr,Wu:2025nwb}. Related calculations have examined Hayward, Horndeski, and Kalb--Ramond backgrounds, infalling sources, and the organization of multiple photon rings \cite{Guo2023,Chen2023,Luo2024Horndeski,TanLan2026KalbRamond,Macedo2026Multiphoton}. These studies establish the optical relevance of the second exterior, but they also show why the metric, shell matching, and emission prescription must be controlled simultaneously.

A separate question is how the removal of a central curvature singularity changes this asymmetric optical system. Regular black-hole geometries demonstrate that nonsingular cores can be obtained by modifying the matter source or the short-distance lapse while retaining a black-hole exterior \cite{AyonBeatoGarcia1998,Hayward2006}. Such regularization is not unique, so its optical consequences must be evaluated for a specified source and under a comparison that does not silently change the asymptotic mass. The present construction uses the new string cloud that extends Letelier's original source by adding a magnetic-like bivector component and yields a nonregular black-hole lapse with a physically restricted string strength \cite{Letelier1979,Alencar2025,Santos2025}. A rational Dagum-type smearing of this source produces finite curvature invariants and an anti--de Sitter core while preserving the string-cloud asymptotics; its horizon sectors, thermodynamics, energy conditions, and shadow scale have been analyzed in the regular seed geometry \cite{Muniz2026}. These properties make it a useful regulator for the present question: it resolves the central singularity, retains a common asymptotic lapse, and supplies an explicit mass shift that can be compensated when regular and nonregular exteriors are compared.

The sensitivity to regularization is not confined to whether an escaping ray reaches the core. In a thin-shell wormhole the core is excised, yet the regularized lapse still changes the retained finite-radius potential, the observer redshift, and the lapse values entering the impact-parameter matching at the throat. Complementary studies of wormhole lensing and thin-shell optical geometry reinforce the broader need to distinguish local propagation from global connectivity \cite{ReboucasEinsteinRings2026,ReboucasPolymerTSW2026}. The specific issue addressed here is therefore whether regularizing neither exterior, only the observer exterior, or both exteriors produces distinguishable changes when the asymptotic masses, throat radius, string-cloud strength, regularization fractions, screen convention, and emission laws are matched.

We address this question by comparing nonregular--nonregular, regular--nonregular and regular--regular asymmetric thin-shell wormholes, together with matched regular and nonregular seed-black-hole controls. We derive the physical impact matching, critical scales, trajectory intervals, orbit numbers and disk transfer functions, and evaluate three phenomenological emission profiles with distinct inner supports \cite{Yue2025}. A single explicitly prescribed circular-plus-plunging emitter model is applied to both polar and inclined views, so brightness changes are not compared across different velocity assumptions. Within the sampled domain, observer-side regularization changes ordinary propagation and shell matching, whereas second-side regularization acts on the returned sector without changing ordinary observer-side rays. The resulting differences depend on source support, emitter motion and viewing angle; the calculation isolates those mechanisms rather than proposing a unique observational signature.

This paper is organized as follows. Section~\ref{sec:geometry} introduces the regular and nonregular string-cloud exteriors, their parameter domains, and the asymptotic-mass matching. Section~\ref{sec:construction} defines the asymmetric shell, the optical domain, and the conserved-quantity matching. Section~\ref{sec:null-geodesics} develops the null potential, photon spheres, critical screen scales, and their regularization dependence. Section~\ref{sec:transfer} analyzes ray classes, representative trajectories, orbit-number families, and transfer functions. Section~\ref{sec:emission} specifies the emission and velocity prescriptions, derives the frequency weights, and compares matched black-hole and wormhole images in face-on and inclined views. Appendix~\ref{app:inclination-comparisons} supplies the complementary inclined comparisons. Section~\ref{sec:conclusion} summarizes the demonstrated results, limitations, and extensions.

\section{Regularized and nonregular string-cloud exteriors}
\label{sec:geometry}

The two sides of the shell are described by static and spherically symmetric geometries in areal coordinates,
\begin{equation}
\begin{aligned}
\dd s_i^2={}&-F_i(r_i)\dd t_i^2+\frac{\dd r_i^2}{F_i(r_i)}+r_i^2\left(\dd\theta^2+\sin^2\theta\,\dd\phi^2\right),
\qquad i=1,2.
\end{aligned}
\label{eq:metric}
\end{equation}
Here $i$ labels the two exterior regions, $t_i$ is the static time coordinate, and $r_i$ is the areal radius, so a symmetry sphere has area $4\pi r_i^2$. The angles $\theta$ and $\phi$ have their standard spherical ranges. The dimensionless lapse $F_i$ must remain positive throughout every retained exterior; this condition fixes the Lorentzian static domain on which the optical construction is performed.

On a regular side, the lapse is the exact rationally regularized string-cloud solution of Ref.~\cite{Muniz2026}. Its source genealogy begins with Letelier's string cloud and its Letelier--Alencar extension \cite{Letelier1979,Alencar2025},
\begin{equation}
\begin{split}
F_{\rm R}(r_i;M_i,\mathcal A,r_{0i})=1-&\left[
\frac{2M_i}{r_i}
-\frac{\mathcal A r_{0i}^{2}}{r_i^{2}}
\Fh\!\left(-\frac12,-\frac14;\frac34;-\frac{r_i^4}{r_{0i}^4}\right)
\right]
\left(1+\frac{r_{0i}}{r_i}\right)^{-4}.
\end{split}
\label{eq:regular-lapse}
\end{equation}
The bare mass $M_i$ and regulator $r_{0i}$ have dimensions of length, while the non-negative string-cloud strength $\mathcal A$ is dimensionless. The regulator satisfies $r_{0i}>0$ on a regular side and controls the radial scale over which the source is smeared. The symbol $\Fh$ denotes the Gaussian hypergeometric function. The source metric admits black-hole, extremal, and horizonless sectors depending jointly on $(M_i,\mathcal A,r_{0i})$; therefore, no independent interval for one of these parameters alone guarantees the required outer horizon.

Near the center, Eq.~\eqref{eq:regular-lapse} gives
\begin{equation}
F_{\rm R}(r_i)=1+\frac{\mathcal A}{r_{0i}^{2}}r_i^2+\mathcal O(r_i^3),
\qquad r_i\rightarrow0.
\label{eq:regular-core}
\end{equation}
For $\mathcal A>0$, this expansion describes a finite-curvature anti--de Sitter core rather than a curvature singularity. The core is not traversed by the rays considered below because the shell removes the entire interval $r_i<R$. Regularity nevertheless changes the exterior lapse and can therefore shift photon spheres, turning points, redshifts, and transfer functions even though the regular center itself is excised.

The large-radius behavior separates the bare mass from the coefficient measured in the string-cloud exterior,
\begin{equation}
F_{\rm R}(r_i)=1-\mathcal A-\frac{2\mu_i}{r_i}
+\mathcal O(r_i^{-4}),
\qquad
\mu_i=M_i-\kappa\mathcal A r_{0i}.
\label{eq:asymptotic-mass}
\end{equation}
The length $\mu_i$ is the coefficient of the leading inverse-radius term, and $\kappa$ is the dimensionless mass-shift constant
\begin{equation}
\kappa=2+\frac{\Gamma(3/4)^2}{\sqrt{\pi}}
=2.847213084793979\ldots,
\label{eq:mass-shift-coefficient}
\end{equation}
where $\Gamma$ is the Euler gamma function. Thus varying $r_{0i}$ at fixed $M_i$ would also vary the leading asymptotic mass. A comparison intended to isolate regularization must instead hold $\mu_i$ fixed and compensate the bare parameter.

The common asymptotic lapse is
\begin{equation}
F_\infty\equiv\lim_{r_i\to\infty}F_i(r_i)=1-\mathcal A>0.
\label{eq:asymptotic-lapse}
\end{equation}
The source requires $\mathcal A\ge0$, while a static two-ended string-cloud exterior requires $\mathcal A<1$. We therefore adopt $0\le\mathcal A<1$. The endpoint $\mathcal A=1$ is excluded because the asymptotic lapse vanishes. Although constant rescalings normalize the temporal and radial sectors, the areal sector retains a solid-angle defect; consequently, $\mu_i$ is not identified with a standard asymptotically flat mass.

Removing the regulator at fixed $(\mu_i,\mathcal A)$ yields the nonregular Letelier--Alencar lapse \cite{Alencar2025}
\begin{equation}
F_{\rm NR}(r_i;\mu_i,\mathcal A)
=1-\mathcal A-\frac{2\mu_i}{r_i}.
\label{eq:nonregular-lapse}
\end{equation}
For $\mu_i>0$ and $0\le\mathcal A<1$, this geometry has one positive horizon and a curvature singularity at the origin. The limit $r_{0i}\to0$ at fixed $M_i$ and $\mathcal A$ recovers Eq.~\eqref{eq:nonregular-lapse}; taking $\mathcal A\to0$ afterward gives Schwarzschild. In contrast, setting $\mathcal A=0$ while keeping $r_{0i}>0$ leaves a rationally regularized Schwarzschild-type geometry. These limits are physically distinct and cannot be interchanged when attributing an optical change to the string cloud or to regularization.

To compare the two lapse families at the same leading asymptotic mass, every regular side uses
\begin{equation}
M_i=\mu_i+\kappa\mathcal A r_{0i}.
\label{eq:bare-mass-matching}
\end{equation}
This matching prescription removes a leading mass shift from the comparison. The remaining changes begin in the regulator-dependent exterior structure and therefore provide a cleaner measure of how the regular geometry alters null propagation.

\section{Asymmetric shell matching and optical domain}
\label{sec:construction}

Following the thin-shell junction construction \cite{Israel1966,Visser1989}, we join two retained exteriors across a static timelike shell,
\begin{equation}
\begin{aligned}
\Mone &: r_1\ge R,\quad F_1(r_1)>0,\\
\Mtwo &: r_2\ge R,\quad F_2(r_2)>0,\\
\Sigma &: r_1=r_2=R.
\end{aligned}
\label{eq:retained-exteriors}
\end{equation}
The manifolds $\Mone$ and $\Mtwo$ are the first and second retained exterior, respectively. Their identified boundary $\Sigma$ is the throat and $R$ is its constant areal radius. Region $\Mone$ contains the observer and emitting disk, whereas $\Mtwo$ supplies the second exterior. The construction requires $R$ to lie strictly outside every retained outer horizon; placing the shell on a simple or degenerate horizon would invalidate the static timelike matching used here.

We use regular (R) and nonregular (NR) for the two lapse choices. The upright scenario label R is distinct from the italic throat radius $R$. The three mass-asymmetric configurations are nonregular--nonregular (NR--NR), R--NR, and R--R, with the first entry always referring to $\Mone$. 
The adopted domain remains $0\le\mathcal A<1$, $\mu_i>0$, and $r_{0i}>0$ on regular sides, supplemented by the root and throat inequalities below. Holding $(\mathcal A,\mu_1,\mu_2,R)$ fixed ensures that the three cases differ by which exterior is regularized rather than by their leading mass or throat scale.

The induced time on the static shell obeys
\begin{equation}
\dd\tau^2=F_i(R)\dd t_i^2,
\label{eq:shell-time}
\end{equation}
where $\tau$ is proper time measured on $\Sigma$. A positive $F_i(R)$ makes the shell time real and is therefore both a geometric and optical admissibility condition.

For a photon crossing the static shell, equality of the locally measured frequency and conservation of angular momentum give
\begin{equation}
\frac{E_1}{\sqrt{F_1(R)}}
=\frac{E_2}{\sqrt{F_2(R)}},
\qquad L_1=L_2\equiv L.
\label{eq:conserved-matching}
\end{equation}
Here $E_i$ is the positive coordinate energy on side $i$, and $L$ is the common angular momentum of the ray. Coordinate energies need not agree because the static times have different normalizations at the shell.

With $b_i\equiv L/E_i$, Eq.~\eqref{eq:conserved-matching} becomes
\begin{equation}
b_1=Zb_2,
\qquad
Z=\sqrt{\frac{F_2(R)}{F_1(R)}}.
\label{eq:impact-matching}
\end{equation}
The length $b_i$ is the coordinate impact parameter on side $i$, and the positive dimensionless factor $Z$ maps second-side impacts to the observer side. It is real only when both throat lapses are positive. Because $Z$ depends on both geometries at $R$, a change confined to $\Mtwo$ can move an observed critical feature even though neither the observer nor the disk leaves $\Mone$.

\section{Null geodesics and critical structure}
\label{sec:null-geodesics}

On the equatorial plane, the radial null equation is
\begin{equation}
\left(\frac{\dd r_i}{\dd\lambda}\right)^2
=E_i^2\left[1-b_i^2V_i(r_i)\right],
\qquad
V_i(r_i)=\frac{F_i(r_i)}{r_i^2}.
\label{eq:null-radial}
\end{equation}
The parameter $\lambda$ is affine and $V_i$ is the null effective potential, with dimensions of inverse length squared. Radial motion is allowed where $1-b_i^2V_i\ge0$ and reverses at a simple zero. Hence regularization can alter a trajectory without changing the matching rule: it changes the shape and extrema of $V_i$ itself.

An unstable circular null orbit and its critical coordinate impact satisfy
\begin{equation}
r_{{\rm ph}i}F_i'(r_{{\rm ph}i})-2F_i(r_{{\rm ph}i})=0,
\qquad
b_{ci}=\frac{r_{{\rm ph}i}}{\sqrt{F_i(r_{{\rm ph}i})}}.
\label{eq:photon-sphere}
\end{equation}
The radius $r_{{\rm ph}i}$ is the selected circular null orbit and $b_{ci}$ is its positive critical impact. A prime denotes differentiation with respect to the corresponding areal radius. Only roots in the retained positive-lapse exterior with $V_i''(r_{{\rm ph}i})<0$ are accepted. This second-derivative condition is essential because a stationary point alone does not identify the unstable orbit that controls the logarithmic enhancement of the deflection angle.

For a static observer in the asymptotic string-cloud region, the physical screen radius is
\begin{equation}
\beta_i=\sqrt{F_\infty}\,b_i.
\label{eq:screen-impact}
\end{equation}
The length $\beta_i$ removes the constant time normalization from the coordinate impact. Plotting $b_i$ directly would therefore mix optical changes with the nonunit asymptotic lapse. At fixed $\mathcal A$, the same $F_\infty$ is used in all three regularization comparisons, so differences in $\beta_i$ trace the finite-radius geometry and shell matching rather than a changing screen convention. When $\mathcal A$ itself is varied below, $F_\infty$ is recomputed for every curve according to Eq.~\eqref{eq:asymptotic-lapse}; the plotted normalization therefore remains physical rather than being frozen at its value at $\mathcal A=0.5$.

The nonregular lapse provides an analytic check,
\begin{equation}
r_{{\rm ph}i}^{\rm NR}=\frac{3\mu_i}{F_\infty},
\qquad
\beta_{ci}^{\rm NR}=\frac{3\sqrt{3}\,\mu_i}{F_\infty}.
\label{eq:nonregular-critical}
\end{equation}
Both scales increase as $\mathcal A$ approaches one at fixed $\mu_i$, reflecting the shrinking asymptotic lapse. This divergence also shows why $\mathcal A=1$ is outside the adopted optical domain. Equation~\eqref{eq:nonregular-critical} is reproduced by the numerical root finder and anchors the regular calculations to a known limit.

The reflection mechanism considered here requires
\begin{equation}
\max(r_{h1},r_{h2})<R
<\min(r_{{\rm ph}1},r_{{\rm ph}2}),
\label{eq:optical-domain}
\end{equation}
where $r_{hi}$ is the largest positive horizon root on side $i$. The lower inequality keeps both retained exteriors static, while the upper inequality allows a ray to cross the shell before reaching either potential maximum. The upper bound is specific to the reflected-ray configuration studied here; it is not a universal requirement for every thin-shell geometry.

\begin{figure}[t]
\centering
\includegraphics[width=0.323\linewidth]{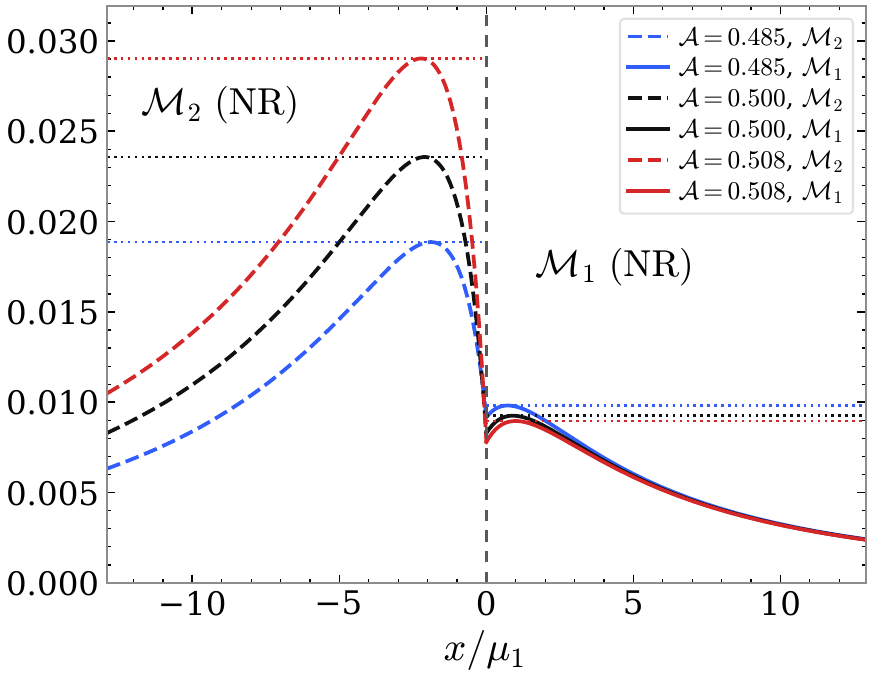}\hfill
\includegraphics[width=0.323\linewidth]{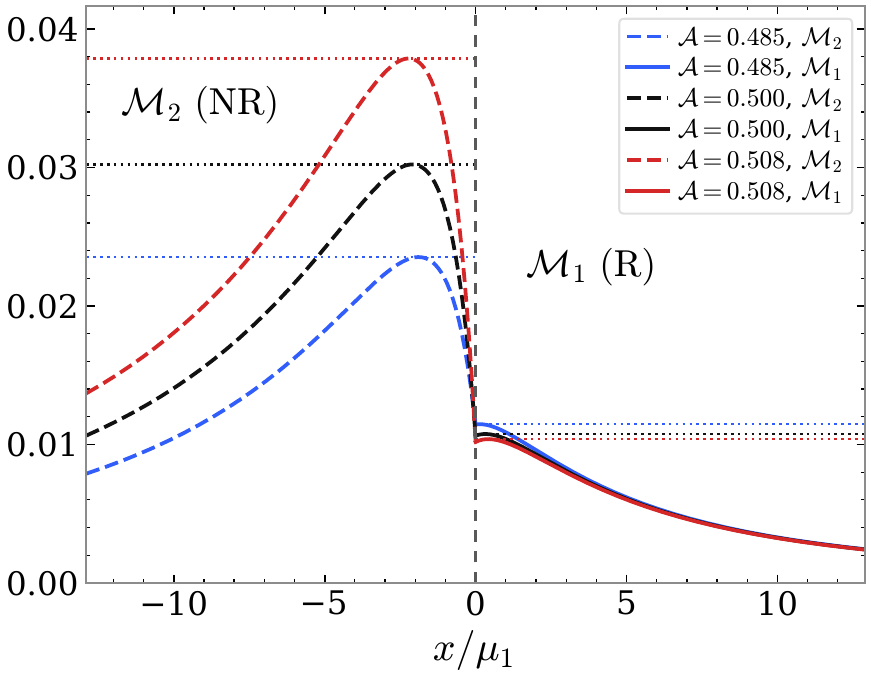}\hfill
\includegraphics[width=0.323\linewidth]{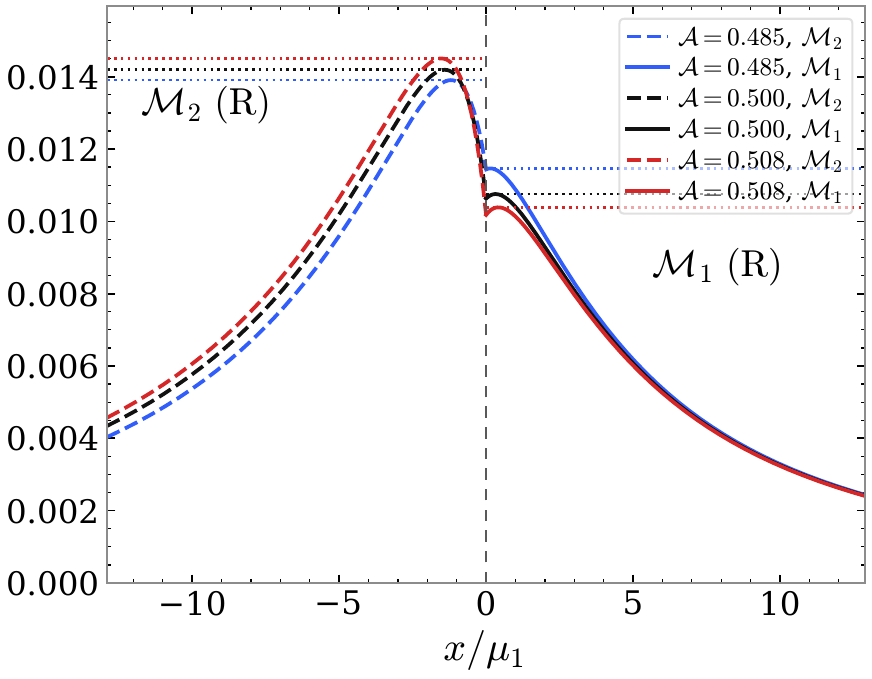}
\caption{Matched null effective-potential profiles for the NR--NR, R--NR, and R--R constructions, respectively, from left to right. The signed coordinate is $x=r_1-R$ on the observer side ($x>0$) and $x=R-r_2$ on the second exterior ($x<0$). The common ordinate within each panel is $\mu_1^2V_1/F_\infty$ on $\Mone$ and $\mu_1^2V_2/(F_\infty Z^2)$ on $\Mtwo$, so the two sides are expressed in the same observer-matched turning-point convention. Blue, black, and red denote $\mathcal A=0.485$, $0.500$, and $0.508$, respectively; solid curves belong to $\Mone$, dashed curves to $\Mtwo$, and dotted segments give the corresponding side-specific critical levels. The remaining parameters are $\mu_1=1$, $\mu_2=1.2$, $R=5.1\mu_1$, and $r_{0i}=0.1\mu_i$ on regular sides. The three panels use independent ordinate ranges; amplitudes across panels must therefore be compared using their numerical tick values rather than their displayed heights.}
\label{fig:potential-comparison}
\end{figure}

To examine the string-cloud dependence without changing the throat or the asymptotic masses, the potential comparison uses $\mathcal A\in\{0.485,0.500,0.508\}$. For the fixed parameters in the caption of Fig.~\ref{fig:potential-comparison}, imposing Eq.~\eqref{eq:optical-domain} simultaneously on the NR--NR, R--NR, and R--R constructions restricts the common fixed-$R$ interval to approximately $0.471688<\mathcal A<0.529412$. The displayed values are deliberately nonuniform: they avoid the rapidly increasing matching sensitivity near the upper boundary while retaining resolvable changes between adjacent profiles. At every sampled $\mathcal A$, $F_\infty$ is updated through Eq.~\eqref{eq:asymptotic-lapse}, and the regular bare masses are recomputed through Eq.~\eqref{eq:bare-mass-matching} so that $(\mu_1,\mu_2,R,r_{01}/\mu_1,r_{02}/\mu_2)$ remain fixed.

Figure~\ref{fig:potential-comparison} compares intrinsic observer-side potentials with second-side potentials expressed in the observer-matched convention. In the NR--NR and R--NR constructions, $\Mtwo$ is the same nonregular geometry at each fixed $\mathcal A$; hence its unscaled potential $V_2$, defined in Eq.~\eqref{eq:null-radial}, is identical in those two cases. The different dashed heights do not represent different intrinsic $\Mtwo$ geometries. They arise because the plotted quantity is $V_2/(F_\infty Z^2)$, and regularizing $\Mone$ changes $F_1(R)$ and therefore the matching factor $Z$ in Eq.~\eqref{eq:impact-matching}. The figure consequently displays how the same second-side potential is mapped to two different observer screens.

The same-color branches meet continuously at the throat in every panel. At $r_i=R$, the factor $Z^{-2}$ maps the second-side potential to the corresponding observer-side value, even though the radial derivatives of the two profiles need not coincide. The resulting cusp records the asymmetric gluing rather than a discontinuity in the photon turning condition. On $\Mone$, increasing $\mathcal A$ produces a modest reduction of the matched potential over the displayed interval. The identical right-hand profiles in the R--NR and R--R panels provide the complementary check: changing only $\Mtwo$ does not alter the intrinsic observer-side potential.

The larger parameter separation occurs in the mapped $\Mtwo$ sector when that side is nonregular. From Eqs.~\eqref{eq:nonregular-lapse} and \eqref{eq:impact-matching}, increasing $\mathcal A$ lowers $F_2(R)$ and enhances the observer-mapped factor $Z^{-2}$, raising both the dashed maximum and its dotted critical level in the NR--NR and R--NR panels. When both sides are regular, the displayed family remains more tightly grouped, indicating a weaker net sensitivity of the matched potential for this particular fixed-$R$ scan. These are numerical trends over the three admissible samples, not global monotonic statements. Moreover, the potential profiles determine turning points and critical trajectories; they do not by themselves determine the brightness of the associated image features, which also depends on the transfer functions and emission prescriptions introduced below.

Table~\ref{tab:critical} reports the accepted roots and screen impacts. All lengths are normalized by $\mu_1$, and $Z\beta_{c2}$ is the second-side critical impact mapped to the observer screen.
\begin{center}
\refstepcounter{table}\label{tab:critical}
\small\textsc{Table \thetable.} Critical structure for $\mathcal A=0.5$, $\mu_2/\mu_1=1.2$, $R/\mu_1=5.1$, and $r_{0i}/\mu_i=0.1$ on regular sides, with $\mu_1$ as the length unit. Where applicable, the regular bare masses are $M_1=1.1423606542\mu_1$ and $M_2=1.3708327851\mu_1$, from Eq.~\eqref{eq:bare-mass-matching}.
\par\medskip
\begin{ruledtabular}
\begin{tabular}{lccccccc}
Case & $r_{h1}$ & $r_{h2}$ & $r_{{\rm ph}1}$ & $r_{{\rm ph}2}$ & $Z$ & $\beta_{c1}$ & $Z\beta_{c2}$\\
\hline
NR--NR & 4.000 & 4.800 & 6.000 & 7.200 & 0.522 & 10.392 & 6.513\\
R--NR  & 3.566 & 4.800 & 5.422 & 7.200 & 0.461 & 9.642  & 5.752\\
R--R   & 3.566 & 4.279 & 5.422 & 6.506 & 0.725 & 9.642  & 8.394\\
\end{tabular}
\end{ruledtabular}
\end{center}

The first two rows isolate regularization of the observer side. Both $r_{h1}$ and $r_{{\rm ph}1}$ move inward, and the observer critical curve decreases from $10.392\mu_1$ to $9.642\mu_1$. The mapped second-side edge also moves inward because $F_1(R)$ changes the matching factor, even though side 2 remains nonregular. Regularizing side 2 in the last row leaves the observer critical curve unchanged but raises the mapped second-side edge to $8.394\mu_1$. The reflected screen interval consequently has widths $3.880\mu_1$, $3.890\mu_1$, and $1.248\mu_1$ in the NR--NR, R--NR, and R--R cases, respectively. Thus observer-side regularization mainly translates this interval for the parameters specified in Table~\ref{tab:critical}, whereas second-side regularization strongly contracts it. This conclusion is local to the sampled parameter set and is not promoted to a global monotonic theorem.

All positive horizons were located by bracketed searches over the full sampled radial domain. Circular-null roots were then filtered by the positive-lapse and negative-curvature conditions following Eq.~\eqref{eq:photon-sphere}. The nonregular analytic values, continuous impact matching, and the three inequalities in Eq.~\eqref{eq:optical-domain} were independently recovered in every case. These checks support the entries in Table~\ref{tab:critical}; they do not establish behavior outside the adopted parameter domain.

\section{Ray topology, orbit numbers, and transfer functions}
\label{sec:transfer}

The two critical impacts divide noncritical rays into three mutually exclusive classes,
\begin{equation}
\begin{cases}
\beta_1>\beta_{c1}, & \text{observer-side scattering},\\
Z\beta_{c2}<\beta_1<\beta_{c1}, & \text{crossing, reflection, and return},\\
0<\beta_1<Z\beta_{c2}, & \text{transmission to the second exterior}.
\end{cases}
\label{eq:ray-classes}
\end{equation}
The reflected family exists only when $Zb_{c2}<b_{c1}$, which holds for all three rows of Table~\ref{tab:critical}. At either equality, the deflection grows without bound as the ray approaches an unstable circular orbit. Such exactly critical trajectories are limiting solutions and are not treated as finite numerical rays.

Introducing the inverse radius $u_i\equiv1/r_i$, the magnitude of the azimuthal quadrature is
\begin{equation}
\frac{\dd\phi}{\dd u_i}
=\frac{1}{\sqrt{b_i^{-2}-u_i^2F_i(1/u_i)}}.
\label{eq:azimuthal-quadrature}
\end{equation}
The variable $u_i$ has dimensions of inverse length. The square-root zeros are turning points, and their integrable endpoint singularities become increasingly sharp near $b_{ci}$. The calculations therefore use endpoint-aware quadrature and dedicated critical-impact refinement rather than a uniform mesh alone.

The impact-space classes in Eq.~\eqref{eq:ray-classes} are first illustrated by representative trajectories before their accumulated winding is organized into orbit-number families.

\begin{figure}[p]
\centering
{\setlength{\tabcolsep}{1.2pt}
\renewcommand{\arraystretch}{0}
\begin{tabular}{@{}ccc@{}}
\includegraphics[scale=0.483]{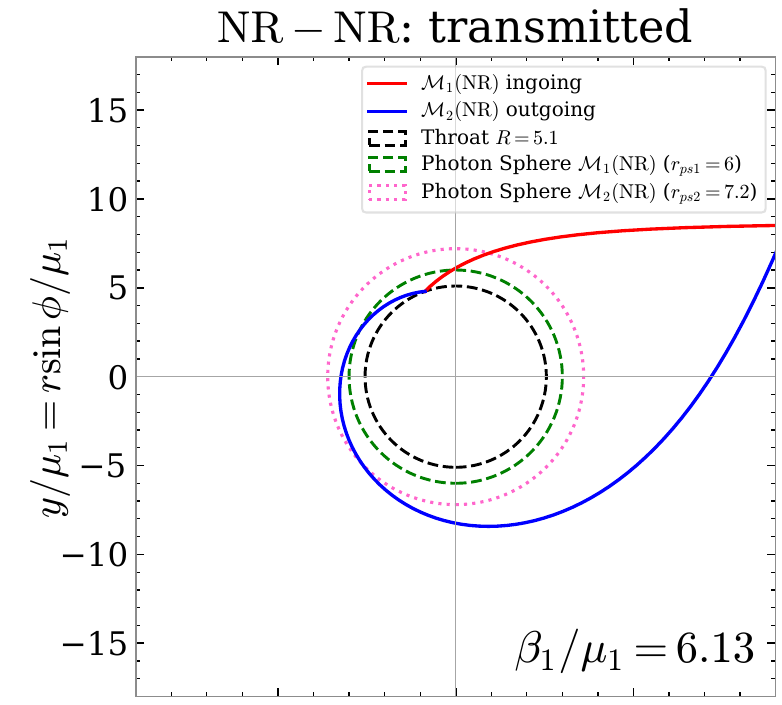} &
\includegraphics[scale=0.483]{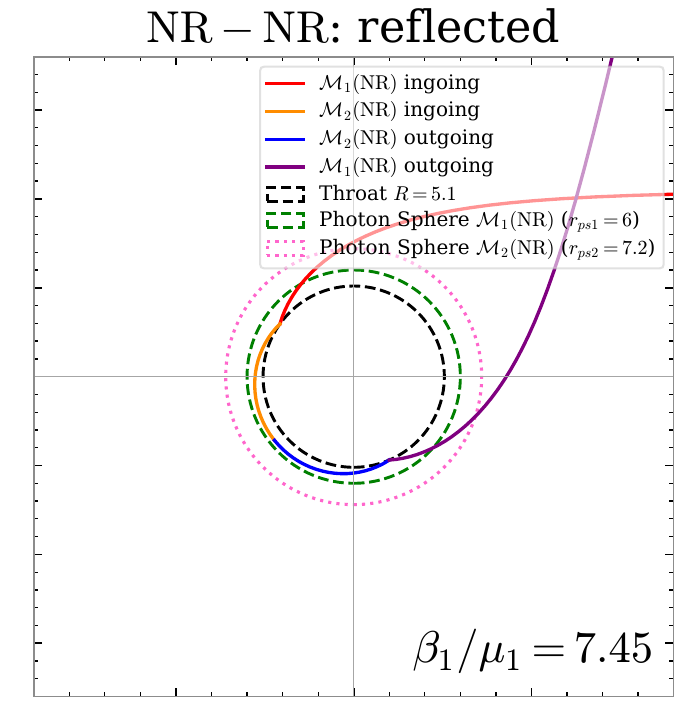} &
\includegraphics[scale=0.483]{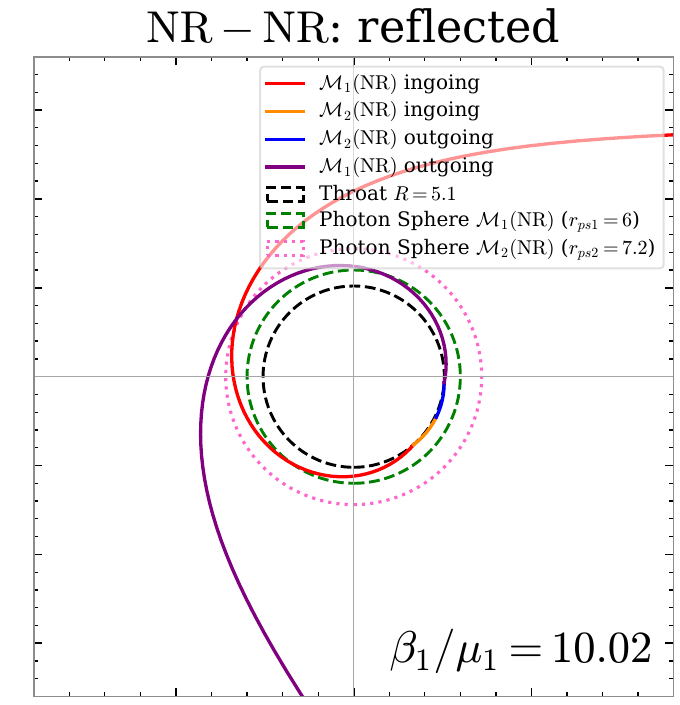} \\[7.8pt]
\includegraphics[scale=0.483]{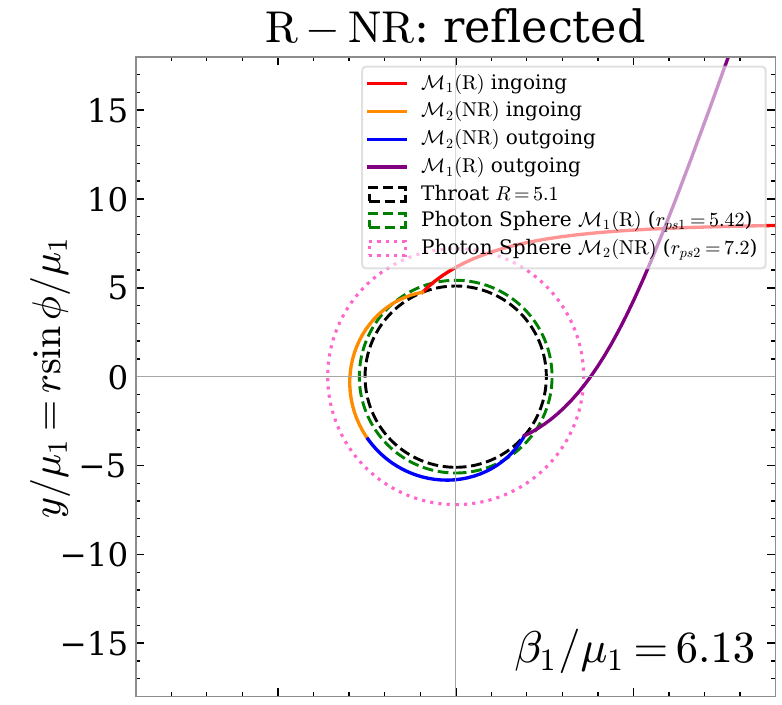} &
\includegraphics[scale=0.483]{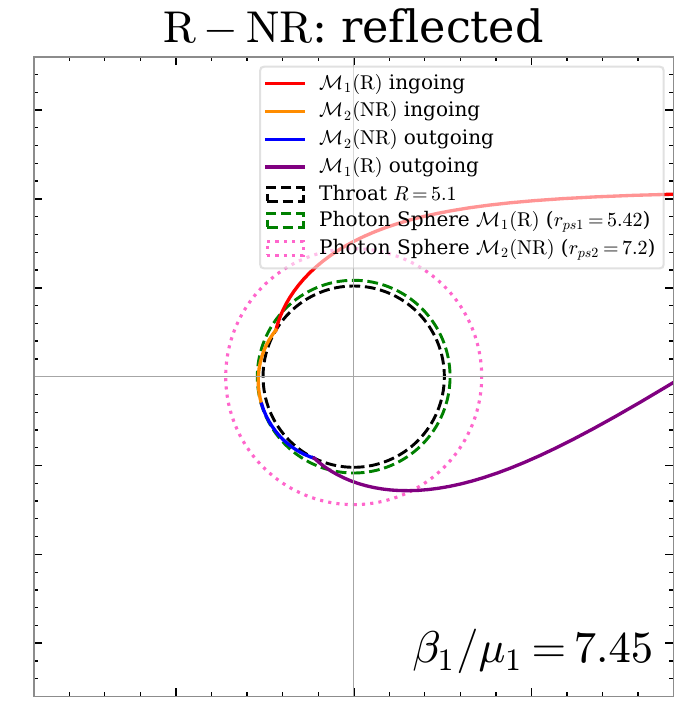} &
\includegraphics[scale=0.483]{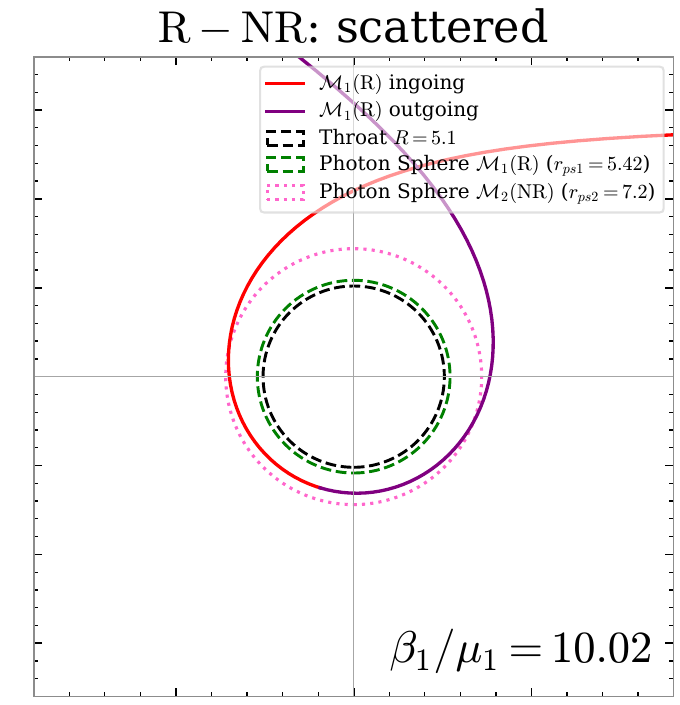} \\[7.8pt]
\includegraphics[scale=0.483]{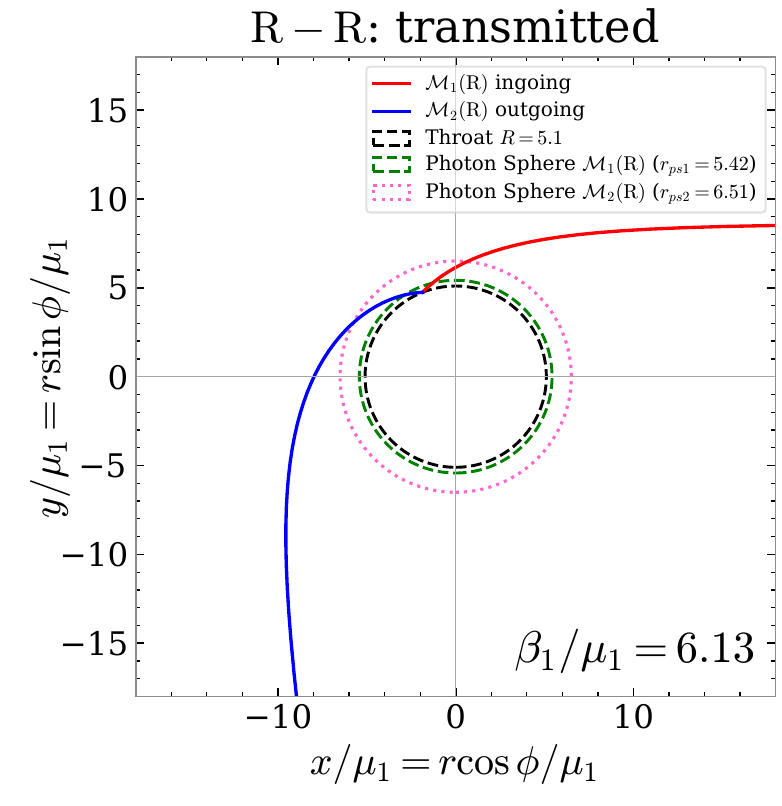} &
\includegraphics[scale=0.483]{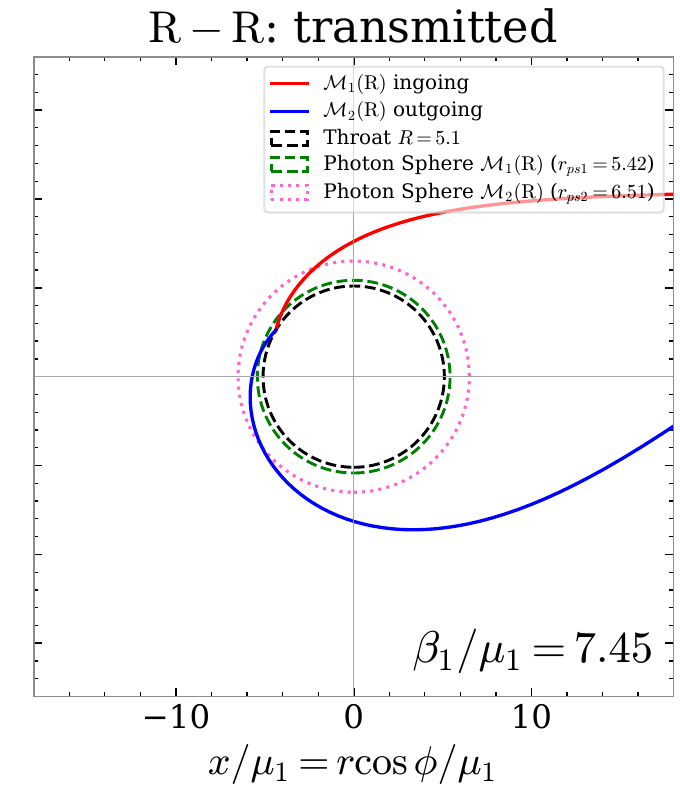} &
\includegraphics[scale=0.483]{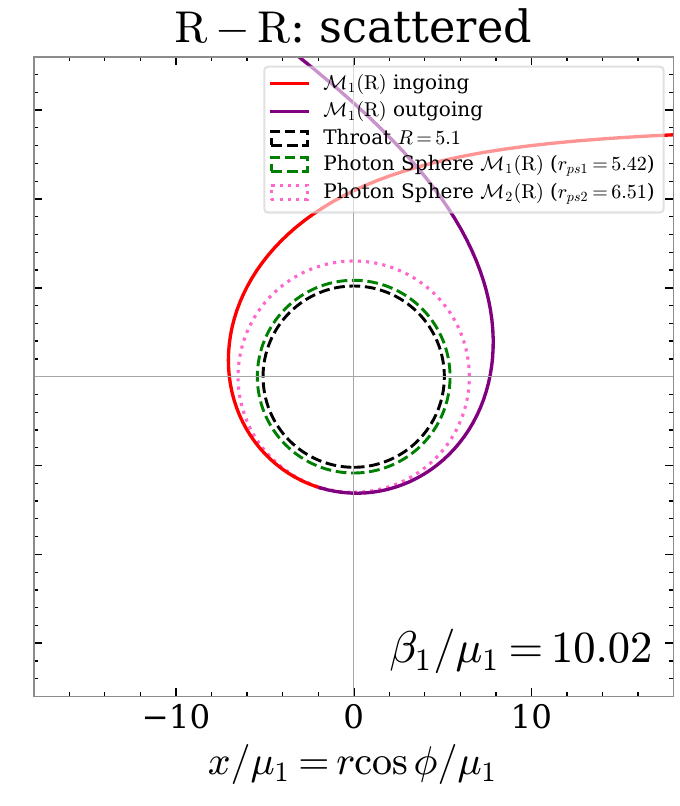}
\end{tabular}}
\caption{Fixed-screen-impact comparison of noncritical null trajectories with $\mathcal A=0.5$, $\mu_2=1.2\mu_1$, $R=5.1\mu_1$, and $r_{0i}=0.1\mu_i$ on regular sides. Rows correspond, from top to bottom, to the NR--NR, R--NR, and R--R constructions, while columns hold the physical observer-screen impact fixed at $\beta_1/\mu_1=6.13$, $7.45$, and $10.02$, respectively. The resulting ray classes are transmitted--reflected--reflected in the top row, reflected--reflected--scattered in the middle row, and transmitted--transmitted--scattered in the bottom row. Hence changes within a column expose a change of ray class caused by the regularization assignment rather than by a retuning of the screen impact. Red and purple segments are ingoing and outgoing branches in $\Mone$, while orange and blue segments are ingoing and outgoing branches in $\Mtwo$, respectively. The black, green, and magenta circles mark the throat, the $\Mone$ photon sphere, and the $\Mtwo$ photon sphere. Each secondary top-axis label identifies the ordered topology and the resulting ray class; the R or NR designation of each exterior is also retained in the legends. Numeric x-axis labels omitted from the upper rows and numeric y-axis labels omitted from the last two columns are represented by the adjacent bottom and left panels, respectively. The auxiliary Cartesian coordinates are normalized by $\mu_1$ and overlay the two radial charts for visualization; they are not a spatial embedding of the wormhole.}
\label{fig:trajectory-grid}
\end{figure}

Figure~\ref{fig:trajectory-grid} applies the intervals in Eq.~\eqref{eq:ray-classes} at common observer-screen impacts. In the first column, $\beta_1/\mu_1=6.13$ lies below the NR--NR mapped edge $Z\beta_{c2}/\mu_1=6.513$, and the ray is transmitted. Regularizing only $\Mone$ lowers that edge to $5.752$, placing the same impact inside the reflected interval of the R--NR construction. This transmitted-to-reflected change occurs even though $\Mtwo$ remains nonregular, because $F_1(R)$ changes the matching factor in Eq.~\eqref{eq:impact-matching}. Regularizing $\Mtwo$ as well raises the mapped edge to $8.394$, so the R--R ray becomes transmitted again. The sequence transmitted--reflected--transmitted therefore isolates the competing effects of observer-side matching and second-side regularization.

The middle column supplies a cleaner test of second-side regularization. At $\beta_1/\mu_1=7.45$, the ray lies between the two critical edges in both NR--NR and R--NR and is consequently reflected. The observer geometry and the screen impact are identical in the R--NR and R--R panels, but regularizing $\Mtwo$ moves $Z\beta_{c2}/\mu_1$ from $5.752$ to $8.394$. The fixed impact then falls below the mapped edge and the R--R ray is transmitted instead of reflected. This transition directly shows that altering the second exterior can change whether a shell-crossing photon returns to the observer or escapes through the opposite asymptotic region.

At $\beta_1/\mu_1=10.02$, the NR--NR ray remains reflected because it lies below $\beta_{c1}/\mu_1=10.392$. Once $\Mone$ is regularized, its critical impact decreases to $9.642\mu_1$, and the same ray becomes observer-side scattered in both R--NR and R--R. These two scattered trajectories coincide because they turn before reaching the throat and never sample $\Mtwo$; the different magenta photon-sphere markers record the unused second-side geometry. The three impacts were chosen inside the relevant transition windows rather than on a critical curve, avoiding the divergent-deflection limit. Finally, the matrix tests only null-geodesic geometry: no emission function or redshift weighting has yet been applied, so a change of ray class is not by itself a prediction of relative image brightness.

The trajectory matrix gives discrete examples at three screen impacts. To characterize the accumulated angular winding continuously across the corresponding impact intervals, we now introduce the orbit numbers. For a trajectory contained in $\Mone$,
\begin{equation}
n_1=\frac{\phi_1}{2\pi},
\label{eq:orbit-number-side1}
\end{equation}
where $\phi_1$ is the accumulated observer-side azimuth and $n_1$ counts complete angular turns. In the usual thin-disk classification, orbit number below $3/4$ corresponds to direct imaging, values between $3/4$ and $5/4$ identify the lensing-ring sector, and values above $5/4$ define the photon-ring sector \cite{Gralla2019}. The boundaries classify trajectory order; they do not by themselves determine whether the associated feature is bright.

For reflected trajectories, the two returned-image counts are \cite{Peng2021}
\begin{equation}
n_2=\frac{\phi_1+\phi_2}{2\pi},
\qquad
n_3=\frac{2\phi_1+\phi_2}{2\pi}.
\label{eq:reflected-orbit-numbers}
\end{equation}
The angle $\phi_2$ is the round-trip azimuth accumulated between the throat and the turning point in $\Mtwo$. Consequently, regularizing $\Mtwo$ affects $n_2$ and $n_3$ through both its potential and the matching factor, while leaving ordinary rays that remain wholly in $\Mone$ unchanged.

\begin{figure}[!ht]
\centering
{\setlength{\tabcolsep}{1.2pt}
\renewcommand{\arraystretch}{0}
\begin{tabular}{@{}ccc@{}}
\includegraphics[scale=0.483]{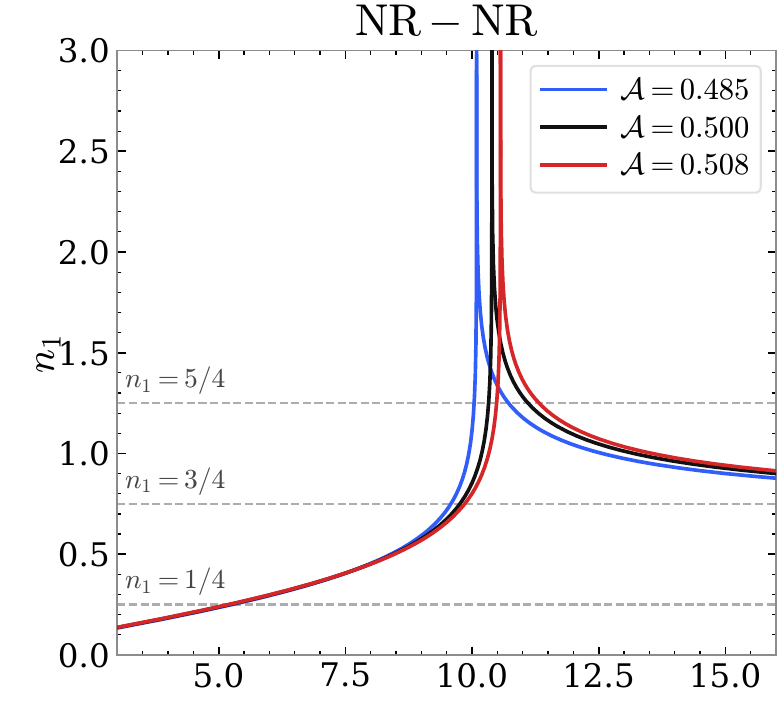} &
\includegraphics[scale=0.483]{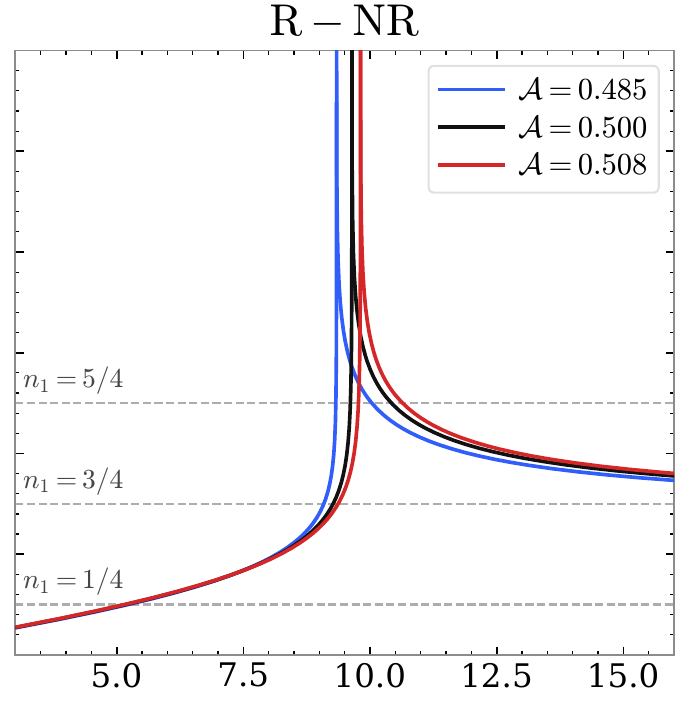} &
\includegraphics[scale=0.483]{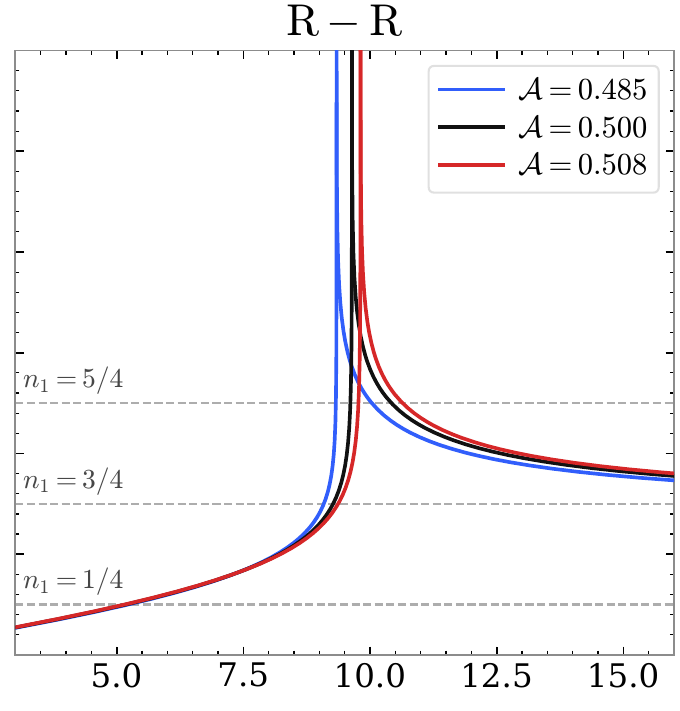} \\[1.6pt]
\includegraphics[scale=0.483]{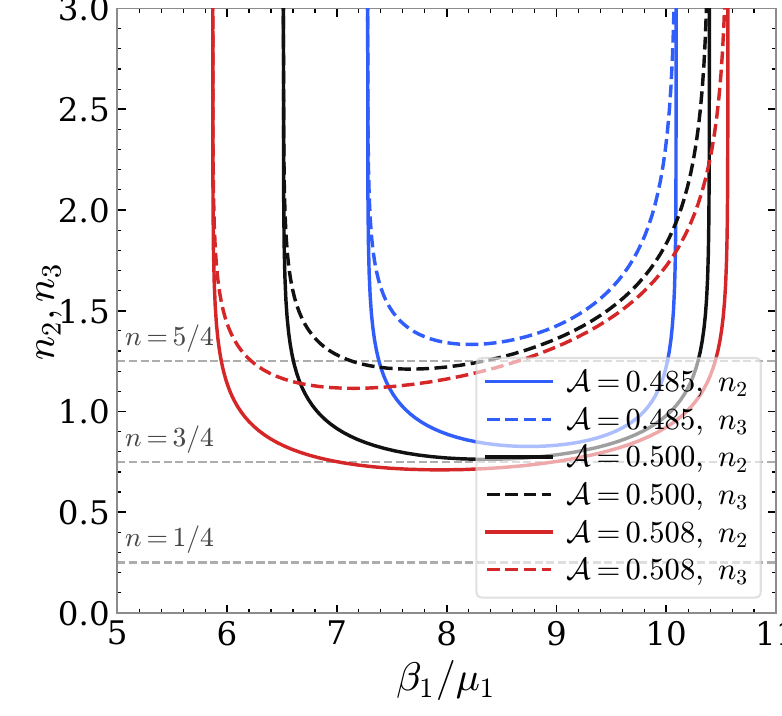} &
\includegraphics[scale=0.483]{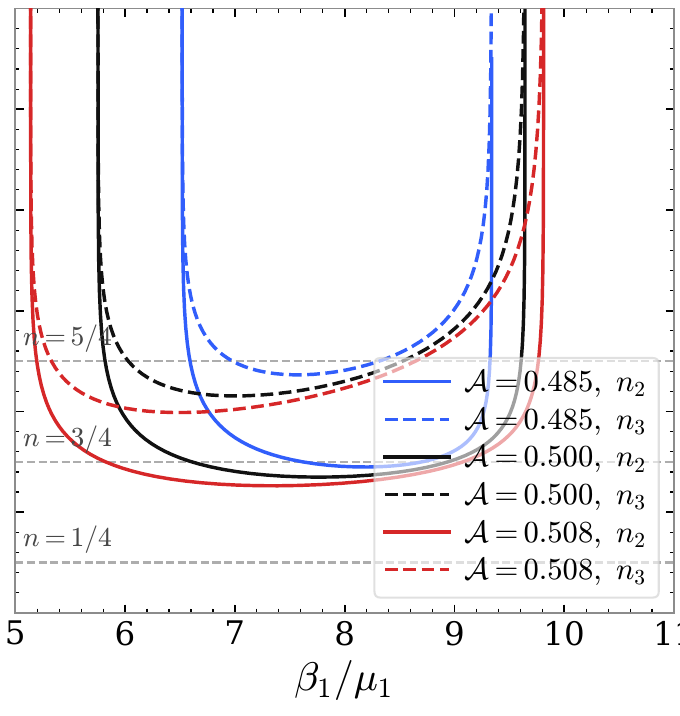} &
\includegraphics[scale=0.483]{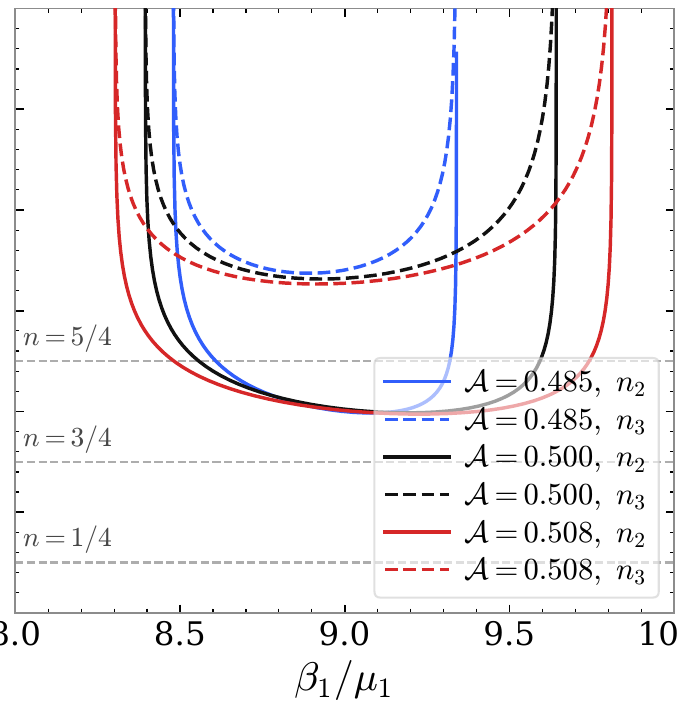}
\end{tabular}}
\caption{Orbit-number families for the three regularization assignments and the admissible string-cloud strengths $\mathcal A=0.485$, $0.500$, and $0.508$. The upper row gives $n_1$ for trajectories confined to $\Mone$, and the lower row gives the two returned-image counts $n_2$ (solid) and $n_3$ (dashed). Columns correspond, from left to right, to NR--NR, R--NR, and R--R, with the observer in $\Mone$. Blue, black, and red identify increasing $\mathcal A$. The gray horizontal levels $n=1/4$, $3/4$, and $5/4$ delimit the direct, lensing-ring, and photon-ring sectors. Numeric y-axis labels omitted from the last two columns are represented by the adjacent left panels. All panels display $0\leq n_i\leq3$; the curves continue above the upper boundary near critical impacts, so this ordinate limit is not a physical cutoff. The upper panels use $3\leq\beta_1/\mu_1\leq16$, the lower NR--NR and R--NR panels use $5\leq\beta_1/\mu_1\leq11$, and the lower R--R panel uses $8\leq\beta_1/\mu_1\leq10$ to resolve its narrower reflected interval. With $\mu_1$ as the length unit, the fixed parameters are $\mu_2/\mu_1=1.2$, $R/\mu_1=5.1$, and $r_{0i}/\mu_i=0.1$ on regular sides. The asymptotic lapse $F_\infty$ and the regular bare masses are recomputed at each $\mathcal A$.}
\label{fig:orbit-number-grid}
\end{figure}

The upper row of Fig.~\ref{fig:orbit-number-grid} isolates the observer-side geometry. Each $n_1$ family develops the expected critical growth at $\beta_{c1}$, with branches on the two sides of the critical impact. Increasing $\mathcal A$ shifts this divergence toward larger $\beta_1/\mu_1$ in both observer-side models over the displayed three-point scan. At equal $\mathcal A$, the critical locations in the two panels with regular $\Mone$ lie to the left of their NR--NR counterparts, consistently with the reduction of $\beta_{c1}/\mu_1$ reported in Table~\ref{tab:critical}. The R--NR and R--R upper panels coincide because these rays turn in $\Mone$ and never probe $\Mtwo$. This equality is therefore a consequence of Eq.~\eqref{eq:orbit-number-side1}, rather than evidence that regularizing the second exterior has no optical effect on shell-crossing trajectories.

The lower row exposes that missing dependence. Reflected rays exist only between the mapped second-side critical impact $Z\beta_{c2}$ and the observer-side critical impact $\beta_{c1}$, which produces the two-ended growth of both $n_2$ and $n_3$. Equation~\eqref{eq:reflected-orbit-numbers} also gives $n_3-n_2=\phi_1/(2\pi)>0$ for these trajectories, explaining why every dashed branch lies above its same-color solid partner. In the NR--NR panel, the minima cross the $3/4$ and $5/4$ levels as $\mathcal A$ varies, so the minimum-order sector itself changes within the sampled family. In R--NR, the boundaries and minima still shift, but all displayed $n_2$ minima remain below $3/4$ and all $n_3$ minima remain between $3/4$ and $5/4$. By contrast, the R--R reflected window is substantially narrower and its curves remain more elevated: throughout the displayed scan, the $n_2$ minima lie in the lensing-ring sector, whereas the $n_3$ minima already exceed $5/4$ and hence belong to the photon-ring sector. The divergence at either edge records asymptotic approach to one of the two unstable photon spheres; it does not represent a finite ray with infinitely many completed orbits.

Taken together, the rows separate two regularization mechanisms. Changing $\Mone$ moves the ordinary critical structure and also modifies every returned branch through $\phi_1$ and the matching factor $Z$. Changing only $\Mtwo$ leaves $n_1$ unchanged but reshapes and contracts the allowed reflected interval through $\phi_2$ and $Z\beta_{c2}$. The continuous families therefore provide the integral counterpart of the fixed-impact class changes shown in Fig.~\ref{fig:trajectory-grid}. The colored curves represent three distinct matched geometries at fixed $(\mu_1,\mu_2,R,r_{01}/\mu_1,r_{02}/\mu_2)$, not the temporal evolution of one wormhole. Moreover, the thresholds determine image order only; whether any corresponding feature is bright or observable still depends on the transfer functions, source support, and redshift weighting introduced below.

\begin{figure}[!ht]
\centering
{\setlength{\tabcolsep}{1.2pt}
\begin{tabular}{@{}ccc@{}}
\includegraphics[scale=0.483]{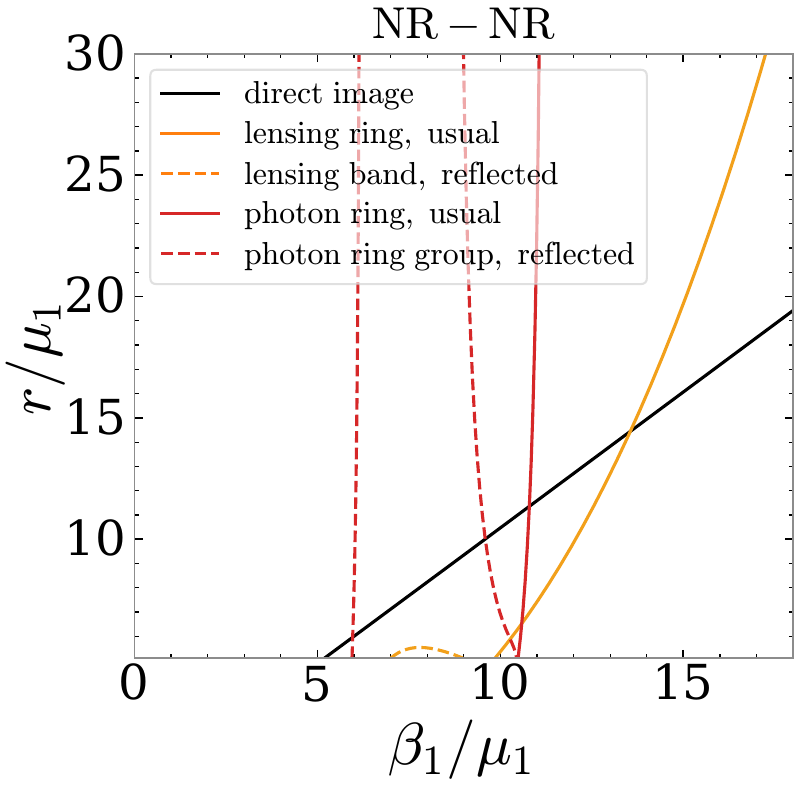} &
\includegraphics[scale=0.483]{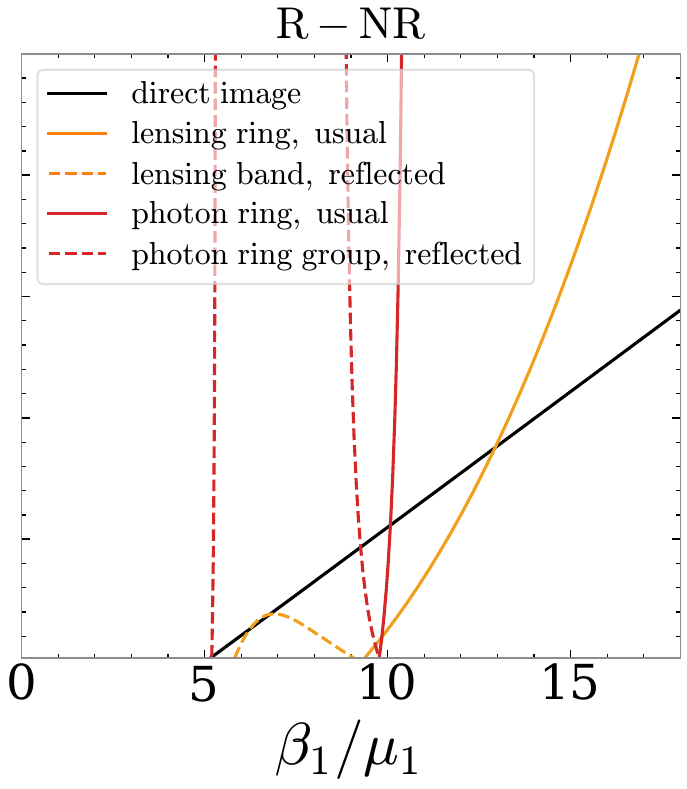} &
\includegraphics[scale=0.483]{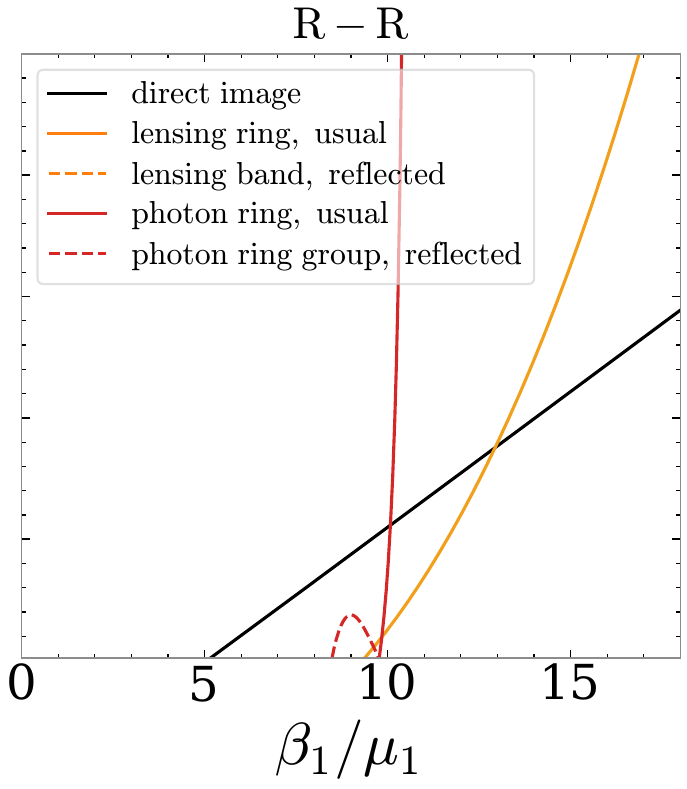}
\end{tabular}}
\caption{Transfer functions at $\mathcal A=0.508$ for the NR--NR, R--NR, and R--R constructions, respectively, from left to right. The ordinate $r_m/\mu_1$ is the radius of the $m$th equatorial-disk intersection and the abscissa $\beta_1/\mu_1$ is the physical observer-screen impact. Numeric y-axis labels omitted from the last two panels are represented by the adjacent left panel. Black, orange, and red solid curves denote the ordinary direct, lensing-ring, and photon-ring branches, while orange and red dashed curves denote reflected second- and third-order branches. The dotted gray level marks the throat at $R=5.1\mu_1$; only intersections with $r_m\ge R$ belong to the retained disk. All panels use the common ranges $0\le\beta_1/\mu_1\le18$ and $5.1\le r_m/\mu_1\le30$. Branches that cross the upper boundary continue to larger emission radii and are not terminated physically at the frame. The remaining parameters are $\mu_1=1$, $\mu_2=1.2$, and $r_{0i}=0.1\mu_i$ on regular sides.}
\label{fig:transfer-comparison}
\end{figure}

For a polar observer and an equatorial geometrically thin disk, intersection order $m$ is fixed by the target angle
\begin{equation}
\phi^{(m)}=\frac{(2m-1)\pi}{2},
\qquad m=1,2,3,\ldots.
\label{eq:intersection-targets}
\end{equation}
The positive integer $m$ labels successive disk crossings. The corresponding transfer function is defined implicitly by
\begin{equation}
\phi\!\left(r_m;\beta_1\right)=\phi^{(m)},
\label{eq:transfer-functions}
\end{equation}
where $r_m(\beta_1)\ge R$ is the emission radius of the $m$th intersection. The first, second, and third ordinary branches generate the direct image, lensing ring, and photon ring, respectively. Reflected rays supply additional second- and third-order branches. Their steep transfer slopes compress broad radial emission intervals into narrow screen intervals, producing strong geometric demagnification even when the source intensity is finite. 
We retain the first three disk-plane targets, $m=1,2,3$, and accept emission only on segments in $\Mone$. A target in the dark second exterior contributes nothing; subsequent observer-side targets keep their original order. Higher orders are not physically absent, and their total omitted intensity is not quantified here.

Figure~\ref{fig:transfer-comparison} first separates effects tied to the observer exterior from those that require shell crossing. The ordinary branches in the R--NR and R--R panels coincide because $\Mone$ is the same regular geometry and these rays reach their disk intersections without probing $\Mtwo$. Their displacement relative to NR--NR is therefore controlled by the observer-side potential and critical impact. This behavior follows directly from Eq.~\eqref{eq:transfer-functions}: changing only $\Mtwo$ cannot alter an angular integral whose entire path remains in $\Mone$.

The dashed branches reveal the complementary second-side dependence. On the 1637-point adaptive impact grid, the reflected second-order branch occupies approximately $7.007\lesssim\beta_1/\mu_1\lesssim8.974$ in NR--NR and $5.826\lesssim\beta_1/\mu_1\lesssim9.074$ in R--NR, whereas no retained reflected second-order solution is found for R--R. The reflected third-order support spans approximately $5.947\lesssim\beta_1/\mu_1\lesssim10.444$, $5.196\lesssim\beta_1/\mu_1\lesssim9.764$, and $8.483\lesssim\beta_1/\mu_1\lesssim9.742$ in NR--NR, R--NR, and R--R, respectively. These endpoints report finite sampled support rather than analytic critical values. Within this scan, regularizing $\Mone$ broadens the reflected second-order interval through both $\phi_1$ and the matching factor, while subsequently regularizing $\Mtwo$ contracts the returned third-order interval and removes the second-order reflected intersection from the retained disk domain.

The nearly vertical portions of the higher-order branches correspond to large $\dd r_m/\dd\beta_1$ and hence strong geometric demagnification: a broad source interval is compressed into a narrow screen interval. The value $\mathcal A=0.508$ is the largest of the three admissible samples used here and was selected because it resolves the reorganization of the reflected branches most clearly. Calculations at $\mathcal A=0.485$ and $0.500$ were retained as numerical controls, but they are not displayed in the manuscript. Neither the appearance nor the widening of a transfer branch implies enhanced brightness; that conclusion requires the emission support and redshift weighting introduced below.

\section{Emission prescriptions and optical observables}
\label{sec:emission}

The observer is static in the asymptotic region of $\Mone$, while the equatorial emitter is an optically and geometrically thin moving disk. The second exterior is nonemitting. The same emitter-motion prescription is used for the face-on and inclined images and for each matched black-hole control; a static spacetime or shell does not imply a static emitter. The observer-side innermost stable circular orbit (ISCO) is the selected outer root of
\begin{equation}
rF_1(r)F_1''(r)+3F_1(r)F_1'(r)
-2rF_1'(r)^2=0.
\label{eq:isco}
\end{equation}
Its radius $r_{\rm ISCO}$ is accepted only where $F_1>0$, $F_1'>0$, and $2F_1-rF_1'>0$. These conditions select a physical timelike circular orbit rather than an arbitrary algebraic root. For $\mathcal A=0.5$, with $r_{01}=0.1\mu_1$ on the regular side, $r_{\rm ISCO}/\mu_1$ decreases from $12.000$ in the NR observer geometry to $10.715$ in the regular observer geometry. This inward displacement changes the formal inner edge of the first emission model together with the geodesic transfer structure.

\subsection{Disk emission and emitter motion}
\label{subsec:thin-disk-emission}

We prescribe an equatorial test flow rather than solve the hydrodynamic equations of accretion. Outside the ISCO, the gas follows circular timelike geodesics; inside it, the gas follows an inward geodesic with the specific energy and angular momentum of the ISCO \cite{He:2025rjq}. Both prescriptions refer to the same observer-side lapse $F_1$. They do not describe a moving observer or a dynamical throat.

With proper time $\tau_{\rm em}$ along the emitter, its four-velocity is $u^\mu=\dd x^\mu/\dd\tau_{\rm em}=(u^t,u^r,0,u^\phi)$. The timelike normalization gives
\begin{equation}
-F_1(u^t)^2+\frac{(u^r)^2}{F_1}+r^2(u^\phi)^2=-1.
\label{eq:emitter-normalization}
\end{equation}
Here $r=r_1$ and $\theta=\pi/2$ on the disk; $u^t$ and $u^r$ are dimensionless, while $u^\phi$ has dimensions of inverse length. The outward radial coordinate fixes $u^r<0$ for inflow.

For $r\ge r_{\rm ISCO}$, the radial geodesic equation with $u^r=0$ yields the coordinate angular velocity
\begin{equation}
\Omega(r)\equiv\frac{u^\phi}{u^t}
=\sqrt{\frac{F_1'(r)}{2r}}.
\label{eq:emitter-angular-velocity}
\end{equation}
The positive root fixes the rotation sense. The quantity $\Omega=\dd\phi/\dd t_1$ has dimensions of inverse length and is not an angular velocity measured with asymptotically normalized proper time. Substituting this relation in Eq.~\eqref{eq:emitter-normalization} gives
\begin{equation}
u^\mu_{\rm circ}(r)
=\frac{(1,0,0,\Omega(r))}
{\sqrt{F_1(r)-\tfrac12 rF_1'(r)}}.
\label{eq:circular-four-velocity}
\end{equation}
The positive square root selects future-directed motion. The conditions following Eq.~\eqref{eq:isco} ensure that the denominator is real and that these are timelike, not null, circular orbits.

\begin{figure}[!ht]
\centering
\includegraphics[width=\linewidth]{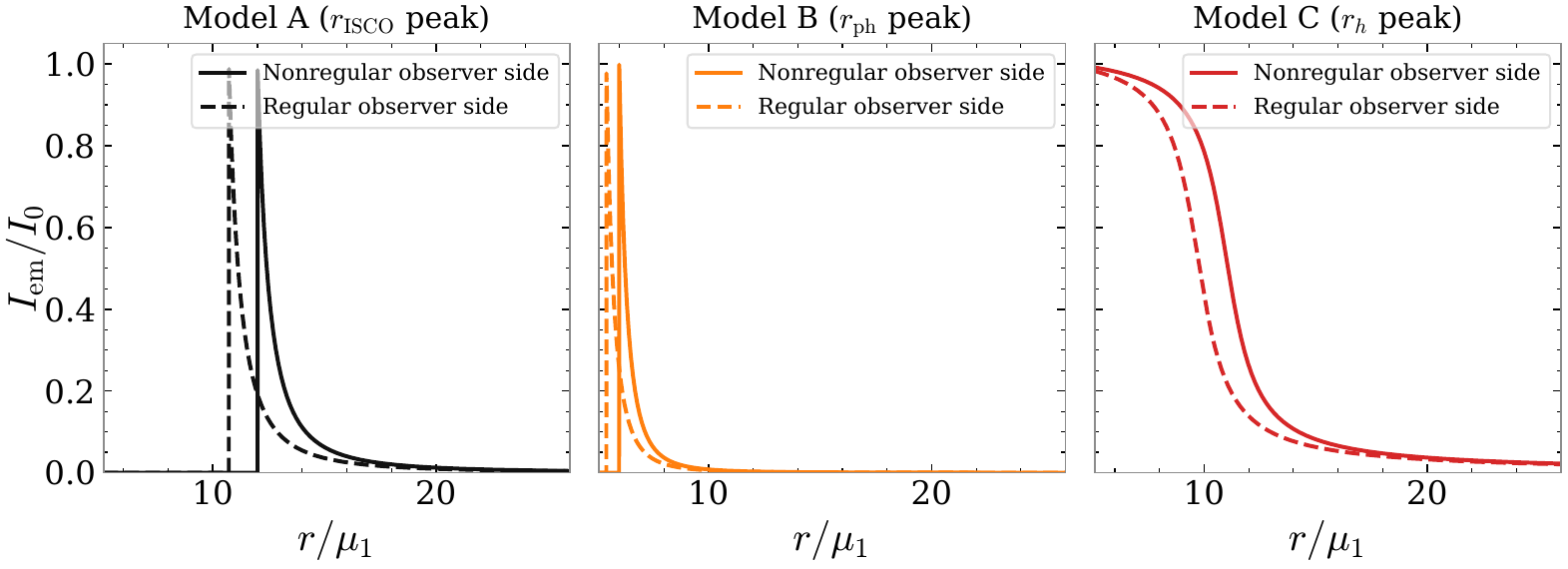}
\caption{Observer-side emission profiles for Models A, B, and C, from left to right, with $\mathcal A=0.5$, $R/\mu_1=5.1$, and $r_{01}/\mu_1=0.1$ on the regular observer side, taking $\mu_1$ as the length unit. Solid curves represent a nonregular observer exterior, while dashed curves represent a regular observer exterior; these two geometries are identified explicitly in each legend. The abscissa is the emission radius $r/\mu_1$, and the common ordinate is the normalized emitted bolometric intensity $I_{\rm em}/I_0$. Model A is anchored at $r_{\rm ISCO}$, Model B at $r_{\rm ph}$, and Model C is formally normalized at the seed horizon $r_h$, as indicated by the secondary upper axes. Only the retained disk domain $5.1\le r/\mu_1\le26$ is displayed. Consequently, the excised seed horizons do not appear in the Model C panel. All panels share the same ordinate range.}
\label{fig:emission-profiles}
\end{figure}

Let $F_{\rm I}\equiv F_1(r_{\rm ISCO})$ and $F'_{\rm I}\equiv F_1'(r_{\rm ISCO})$. The conserved specific energy $\mathcal E_{\rm I}$ and specific angular momentum $\ell_{\rm I}$ of the inward flow are fixed by the circular orbit at the ISCO:
\begin{equation}
\mathcal E_{\rm I}
=\frac{F_{\rm I}}{\sqrt{F_{\rm I}-r_{\rm ISCO}F'_{\rm I}/2}},
\qquad
\ell_{\rm I}
=\frac{r_{\rm ISCO}^{2}\sqrt{F'_{\rm I}/(2r_{\rm ISCO})}}
{\sqrt{F_{\rm I}-r_{\rm ISCO}F'_{\rm I}/2}}.
\label{eq:isco-flow-constants}
\end{equation}
The dimensionless $\mathcal E_{\rm I}=-u_t$ and length $\ell_{\rm I}=u_\phi$ are per unit rest mass and must not be confused with the photon constants $E_i$ and $L$. Killing-energy conservation and Eq.~\eqref{eq:emitter-normalization} then determine the inner prescription,
\begin{equation}
u^\mu_{\rm pl}(r)
=\left(
\frac{\mathcal E_{\rm I}}{F_1(r)},
-\sqrt{\mathcal E_{\rm I}^{\,2}
-F_1(r)\left(1+\frac{\ell_{\rm I}^{\,2}}{r^2}\right)},
0,\frac{\ell_{\rm I}}{r^2}
\right).
\label{eq:plunging-four-velocity}
\end{equation}
This branch is used only for $r<r_{\rm ISCO}$ where $F_1>0$ and the radicand is non-negative. Its velocity approaches the circular ISCO velocity continuously. The exactly marginal circular orbit would remain there without a perturbation; Eq.~\eqref{eq:plunging-four-velocity} specifies the inward separatrix below it, not a finite-time departure from that exact orbit. In a black-hole control the emitting domain extends outside $r_h$; in a wormhole it is restricted to $r\ge R$. No continuation or emission of this flow in $\Mtwo$ is assumed. This geodesic prescription omits pressure, magnetic stresses and time-dependent accretion, so its Doppler weights are model assumptions rather than a unique prediction of the seed metric.

We use three dimensionless phenomenological source profiles from Ref.~\cite{Yue2025}. Model A begins at the ISCO,
\begin{equation}
\frac{I_{\rm em}^{(A)}(r)}{I_0}=
\begin{cases}
\left[\dfrac{\mu_1}{r-(r_{\rm ISCO}-\mu_1)}\right]^2,
& r>r_{\rm ISCO},\\[4pt]
0, & r\le r_{\rm ISCO}.
\end{cases}
\label{eq:emission-a}
\end{equation}
Here $I_{\rm em}^{(A)}$ is the emitted bolometric intensity and $I_0>0$ is the common intensity scale. The profile approaches $I_0$ immediately outside its inner edge and then decays quadratically. Its sharp truncation suppresses every transfer branch whose intersection lies below $r_{\rm ISCO}$; this is an illumination choice, not the removal of the underlying geodesic branch.

Model B is more radially concentrated and starts at the observer-side photon sphere,
\begin{equation}
\frac{I_{\rm em}^{(B)}(r)}{I_0}=
\begin{cases}
\left[\dfrac{\mu_1}{r-(r_{\rm ph}-\mu_1)}\right]^3,
& r>r_{\rm ph},\\[4pt]
0, & r\le r_{\rm ph}.
\end{cases}
\label{eq:emission-b}
\end{equation}
The quantity $I_{\rm em}^{(B)}$ is the second emitted-intensity profile, and $r_{\rm ph}\equiv r_{{\rm ph}1}$. Its cubic decay gives greater relative weight to intersections near the critical region than Model A. The discontinuous cutoff again belongs to the prescribed source and must not be interpreted as a discontinuity of the spacetime.

Model C has formal support down to the outer horizon of the unexcised observer-side seed,
\begin{equation}
\frac{I_{\rm em}^{(C)}(r)}{I_0}=
\begin{cases}
\dfrac{\frac{\pi}{2}-\arctan\!\left[\frac{r-(r_{\rm ISCO}-\mu_1)}{\mu_1}\right]}
{\frac{\pi}{2}-\arctan\!\left[\frac{r_h-(r_{\rm ISCO}-\mu_1)}{\mu_1}\right]},
& r>r_h,\\[8pt]
0, & r\le r_h.
\end{cases}
\label{eq:emission-c}
\end{equation}
The function $I_{\rm em}^{(C)}$ is normalized by the observer-side seed horizon $r_h\equiv r_{h1}$. In the joined spacetime, however, that horizon is excised and the physical disk support is the intersection of the formal source domain with $r\ge R$. Model C therefore begins at the throat rather than at an accessible horizon. Unlike Models A and B, it supplies nonzero emission throughout the entire retained inner disk and can illuminate transfer branches that intersect below the photon sphere or the ISCO.

Figure~\ref{fig:emission-profiles} isolates the illumination change produced by regularizing the observer exterior. Equations~\eqref{eq:emission-a}--\eqref{eq:emission-c} contain only observer-side characteristic radii, so changing \emph{only} $\Mtwo$ cannot alter these source profiles. The solid curves therefore represent the NR--NR construction, whereas the same dashed curves apply to both R--NR and R--R. This degeneracy is exact at fixed observer-side parameters and differs from the returned geodesic branches, which remain sensitive to the second exterior through the matching factor and the $\Mtwo$ angular integral.

Regularization moves the Model A onset inward from $r_{\rm ISCO}/\mu_1=12.000$ to $10.715$ and the Model B onset from $r_{\rm ph}/\mu_1=6.000$ to $5.422$. The latter regular cutoff consequently lies only $0.322\mu_1$ outside the throat. Model C remains nonzero throughout the retained disk in both cases; within the displayed inner region, its regular-observer profile lies below the nonregular one because both $r_{\rm ISCO}$ and the formal normalization radius $r_h$ change in Eq.~\eqref{eq:emission-c}. These differences modify the radial support and weight assigned to disk intersections, but they cannot move the geometric transfer branches of Fig.~\ref{fig:transfer-comparison}. Their screen imprint arises only after the source profile is evaluated at those transfer radii and combined with the redshift, as specified below.

\subsection{Imaging method}\label{subsec:imaging-method}

\begin{figure}[!htb]
\centering
\includegraphics[width=0.66\linewidth]{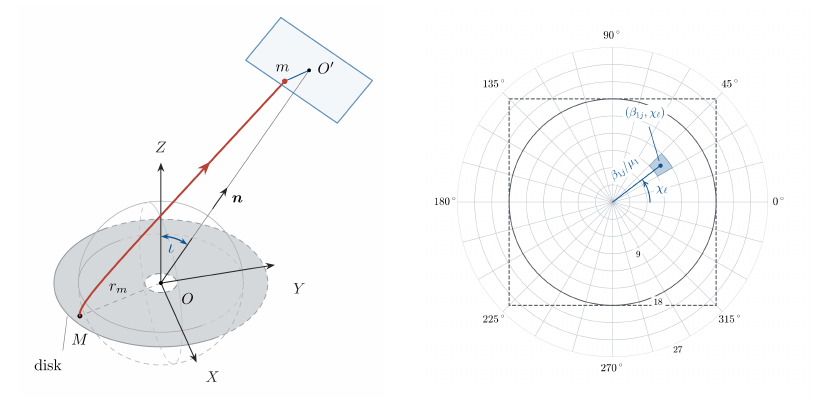}
\caption{Viewing geometry and screen coordinates. Left: the geometric construction used in thin-disk imaging \cite{Luminet:1979nyg,Cui:2026cfx,Sharipov:2025yfw}. The observer direction $\boldsymbol n$ makes an angle $\iota$ with the disk normal, and a representative ray connects the disk point $M$ to the image point $m$. Right: the polar screen grid $(\beta_1,\chi)$, with a shaded cell centered at $(\beta_{1j},\chi_\ell)$. The solid circle and dashed square mark the radial profile window and the displayed image field, respectively.}
\label{fig:screen-polar-grid}
\end{figure}

Owing to spherical symmetry, the null geodesics obtained above can be rotated in an auxiliary Cartesian frame to describe different observer inclinations and positions on the image plane. The viewing geometry is illustrated in the left panel of Fig.~\ref{fig:screen-polar-grid}: the disk lies in the $XY$ plane, and the observer direction $\boldsymbol n$ makes an angle $\iota$ with the disk normal along the $Z$ axis. We consider $0\le\iota<\pi/2$, excluding the exactly edge-on limit for the razor-thin disk. The image plane is perpendicular to $\boldsymbol n$ and is parametrized by the physical screen radius $\beta_1$ and azimuth $\chi\in[0,2\pi)$. As shown in the right panel, each screen cell is indexed by $(\beta_1,\chi)$, with $\chi$ increasing counterclockwise from the rightward screen direction. Inclined intensities are evaluated at native polar-cell centers before interpolation for display. The radial profile window is $0\le\beta_1/\mu_1\le18$, while traced rays extend to $\beta_1/\mu_1=27$ to enclose the displayed square field of side $36\mu_1$. The face-on profiles use the adaptive critical-impact and source-edge mesh specified in Sec.~\ref{subsec:joined-face-on}.

 The required transformation consists of a rotation about the fixed $X$ axis followed by a rotation about the fixed $Y$ axis. In terms of the corresponding generators, the rotation matrix is
\begin{align}
\mathrm{R}(\iota,\chi)
&=e^{(\iota-\pi/2)J_y}e^{\chi J_x}
\nonumber\\
&=
\begin{pmatrix}
\sin\iota & -\cos\iota\sin\chi & -\cos\iota\cos\chi\\
0 & \cos\chi & -\sin\chi\\
\cos\iota & \sin\iota\sin\chi & \sin\iota\cos\chi
\end{pmatrix}.
\label{eq:imaging-rotation}
\end{align}

For each monotonic radial segment $s$ of a noncritical ray at fixed $\beta_1$, let $\phi_{i,s}(r;\beta_1)$ denote the cumulative orbit angle, measured from the observer along the backward-traced ray, in the reference equatorial plane. The quadrature in Eq.~\eqref{eq:azimuthal-quadrature} is evaluated branch by branch in $u_i$. Using $r$ as the parameter, the rotated segment at a specified inclination and image-plane azimuth is
\begin{equation}
\mathrm{R}\boldsymbol{X}_{i,s}(r)
=r
\begin{pmatrix}
\sin\iota\cos\phi_{i,s}(r;\beta_1)-\cos\iota\sin\chi\sin\phi_{i,s}(r;\beta_1)\\
\cos\chi\sin\phi_{i,s}(r;\beta_1)\\
\cos\iota\cos\phi_{i,s}(r;\beta_1)+\sin\iota\sin\chi\sin\phi_{i,s}(r;\beta_1)
\end{pmatrix}.
\label{eq:rotated-trajectory}
\end{equation}
Here $\boldsymbol{X}_{i,s}$ is the geodesic position in the auxiliary Cartesian frame. 
Intersections with the equatorial disk are obtained by setting the third rotated Cartesian component, denoted $\mathsf Z$, to zero. This auxiliary coordinate is distinct from the throat-matching factor $Z$ in Eq.~\eqref{eq:impact-matching}. With the principal branch of the arctangent, the first crossing angle is
\begin{equation}
\phi_c=\frac{\pi}{2}+\arctan(\tan\iota\sin\chi),
\qquad
\frac{\pi}{2}-\iota\leq\phi_c\leq\frac{\pi}{2}+\iota,
\label{eq:first-disk-crossing}
\end{equation}
and the subsequent disk crossings occur at
\begin{equation}
\phi_m=\phi_c+(m-1)\pi,
\qquad m=1,2,\ldots .
\label{eq:successive-disk-crossings}
\end{equation}
For $\iota=0$, Eq.~\eqref{eq:first-disk-crossing} gives $\phi_c=\pi/2$, and Eq.~\eqref{eq:successive-disk-crossings} recovers the polar targets in Eq.~\eqref{eq:intersection-targets}.
The radius of the $m$th accepted intersection is determined implicitly on an observer-side segment by
\begin{equation}
\phi_m=\phi_{1,s}(r_m;\beta_1),\qquad r_m\in\mathcal I_1,\quad r_m\ge R,
\label{eq:inclined-intersection-radius}
\end{equation}
where
\begin{equation*}
\phi_{1,s}(r;\beta_1)=
\begin{cases}
\phi_{1,{\rm in}}(r;\beta_1), & s={\rm in},\\
\phi_{1,{\rm out}}(r;\beta_1)=\phi_{\rm tot}-\phi_{1,{\rm in}}(r;\beta_1), & s={\rm out}.
\end{cases}
\end{equation*}
Both functions measure the cumulative orbit angle from observer-side infinity to radius $r$ in $\Mone$. The labels in and out denote the initial inward and subsequent outward branches of the backward trace, respectively. The total angle $\phi_{\rm tot}$ includes the excursion in $\Mtwo$ for a throat-reflected ray. The common radial interval is $\mathcal I_1=[R,\infty)$ for rays crossing the throat and $\mathcal I_1=[r_{\min},\infty)$ for observer-side scattered rays, where $r_{\min}$ is their turning radius. A ray transmitted to infinity in $\Mtwo$ has only the incoming observer-side branch.

The branch is selected by the angular location of the target $\phi_m$. The incoming branch is tested first; if the target is outside its angular range, the outgoing branch is used only when it exists and contains that target. Numerically, the incoming table is inverted at $\phi_m$ for an incoming intersection and at $\phi_{\rm tot}-\phi_m$ for an outgoing intersection. For a throat-reflected ray, the incoming range is $0<\phi_m\le\phi_{1,{\rm in}}(R;\beta_1)$, and the outgoing range is $\phi_{\rm tot}-\phi_{1,{\rm in}}(R;\beta_1)\le\phi_m<\phi_{\rm tot}$. For an observer-side scattered ray, $R$ in these bounds is replaced by $r_{\min}$. Targets in $\Mtwo$ or beyond the ray yield no observer-side disk intersection.

Equation~\eqref{eq:inclined-intersection-radius} uses the retained wormhole domain. For a seed-black-hole control, the same angular intersection condition is solved with $r_m>r_h$ in place of $r_m\ge R$, and a captured segment ends at the horizon. The frequency prescription is unchanged; only the available path and inner source domain differ.

To evaluate the redshift, we rotate the photon momentum together with the ray. With the normalization $E=-k_t=1$, the standard equatorial covector is
\begin{equation}
k_\mu^{\rm st}=(-1,k_r,0,b),
\label{eq:standard-photon-covector}
\end{equation}
where $E>0$ fixes the future-directed time orientation and the signed impact parameter is $b=-b_1=-\beta_1/\sqrt{F_\infty}$ for the adopted backward trace. A spatial rotation leaves $k_r$ unchanged, while the null condition and the rotation give the two components required at a disk crossing:
\begin{equation}
k_r
=\frac{s_r}{F_1(r)}
\sqrt{1-\frac{F_1(r)b_1^2}{r^2}},
\qquad
k_\phi
=-b_1\sin\iota\cos\chi
=-\frac{\beta_1\sin\iota\cos\chi}{\sqrt{F_\infty}},
\label{eq:photon-momentum-at-disk}
\end{equation}
Here $s_r=\operatorname{sgn}(k^r)$ is fixed by the radial segment: $s_r=+1$ for a future-directed photon moving toward increasing $r$ and $s_r=-1$ for one moving toward decreasing $r$. Because the ray is traced backward from the observer, these signs correspond respectively to the initial and subsequent observer-side segments of a scattered or throat-reflected ray.

For the static asymptotic observer, $u^\mu_{\rm obs}=(F_\infty^{-1/2},0,0,0)$. The measured photon frequency is proportional to $-k_\mu u^\mu$, so the observer-to-emitter frequency ratio at the $m$th accepted crossing is
\begin{equation}
g_m
=\frac{1}{\sqrt{F_\infty}
\left[u^t-u^r k_r-u^\phi k_\phi\right]_{r=r_m}}.
\label{eq:redshift}
\end{equation}
Here the circular or plunging four-velocity from Eqs.~\eqref{eq:circular-four-velocity} and \eqref{eq:plunging-four-velocity} is selected according to the value of $r_m$. Treating the geometrically thin disk as optically thin, the observed bolometric intensity is the sum of the contributions from all accepted crossings,
\begin{equation}
\mathscr I_\iota^{(k)}(\beta_1,\chi)
=\sum_{m\in\mathcal S_\iota(\beta_1,\chi)}
g_m^4 I_{\rm em}^{(k)}(r_m),
\qquad k\in\{A,B,C\},
\label{eq:observed-intensity}
\end{equation}

Here $g_m\equiv\nu_{\rm obs}/\nu_{{\rm em},m}$ is dimensionless, $\nu_{\rm obs}$ and $\nu_{{\rm em},m}$ are the frequencies measured by the observer and the $m$th emitter, and $\mathcal S_\iota(\beta_1,\chi)$ contains the accepted observer-side intersections among the first three disk-plane targets. Emission is isotropic in the local emitter frame; absorption, scattering and emission in $\Mtwo$ are neglected. The invariance of specific intensity divided by the cube of frequency gives a factor $g_m^3$ spectrally, and $\dd\nu_{\rm obs}=g_m\dd\nu_{{\rm em},m}$ supplies the fourth factor after bolometric integration.
For the polar view, axial symmetry reduces Eq.~\eqref{eq:observed-intensity} to
\begin{equation}
\mathscr I_0^{(k)}(\beta_1,\chi)
=I_{\rm obs}^{(k)}(\beta_1),
\label{eq:image-map}
\end{equation}

The image assigned to a polar-screen cell is $\mathscr I_\iota^{(k)}(\beta_1,\chi)$ from Eq.~\eqref{eq:observed-intensity}. In the face-on limit, axial symmetry makes it independent of $\chi$, so the two-dimensional image is obtained from the same radial intensity profile in Eq.~\eqref{eq:image-map}. The unlabelled square maps and quadrant zooms are display windows of this polar-screen calculation.

For $\iota=0$ the azimuthal photon momentum vanishes, but the emitter is still moving. Writing $g_{\rm circ,0}$ for the frequency ratio on a circular emitting branch in this polar view,
\begin{equation}
g_{\rm circ,0}^{\,4}(r)
=\left[\frac{F_1(r)-\tfrac12 rF_1'(r)}{F_\infty}\right]^2.
\label{eq:polar-circular-weight}
\end{equation}
This transverse-Doppler and gravitational weight differs from the static-emitter factor $[F_1(r)/F_\infty]^2$. In the plunging region, the full expression in Eq.~\eqref{eq:redshift} is required even face-on: the term $-u^r k_r$ depends on whether the future-directed photon is locally outgoing or ingoing. Thus identical emission radii can have different frequency weights on different radial segments. A radial intensity curve is not obtained by applying one radius-only static factor to every crossing.

\subsection{Imaging results}\label{subsec:imaging-results}

\subsubsection{Moving-emitter black-hole and thin-shell comparisons}
\label{subsec:joined-face-on}

\begin{figure}[!ht]
\centering
\includegraphics[width=\linewidth,height=0.79\textheight,keepaspectratio]{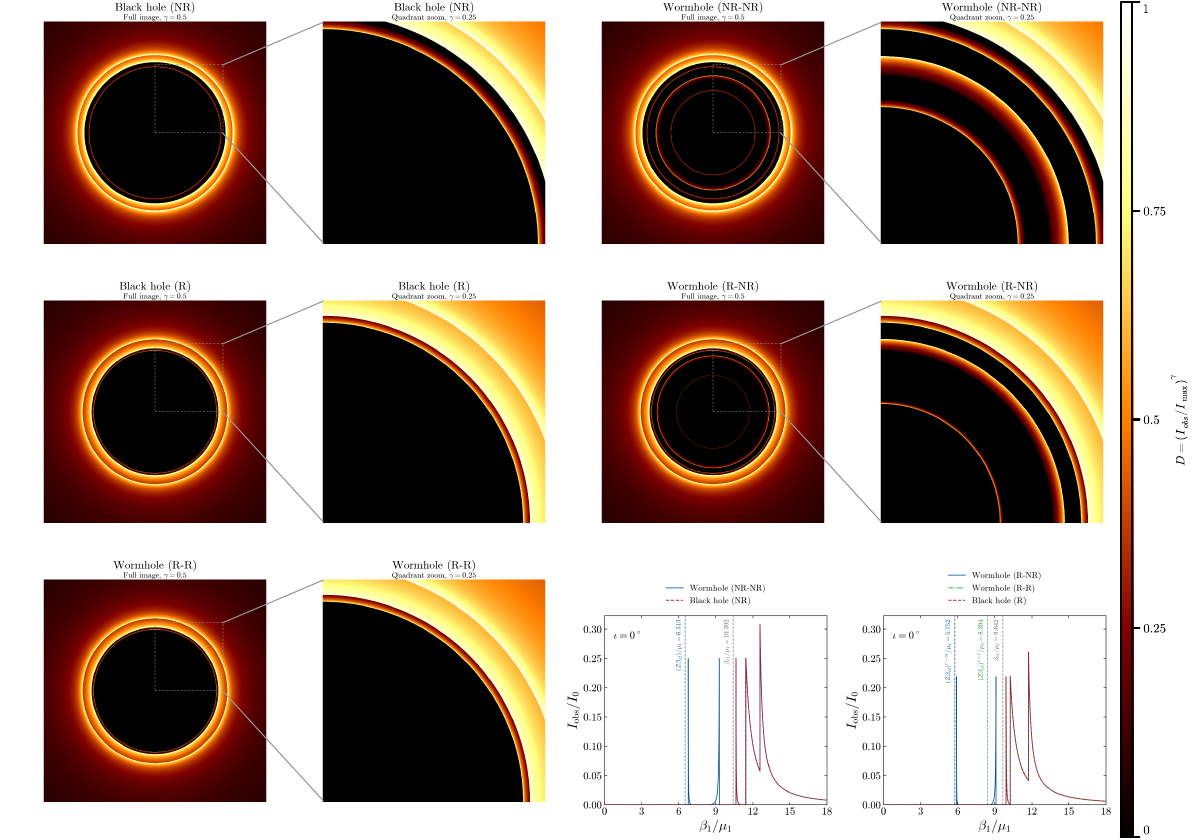}
\caption{Face-on moving-emitter observables for Model A for wormhole parameters $\mathcal A=0.5$, $\mu_2/\mu_1=1.2$, $R/\mu_1=5.1$, and $r_{0i}/\mu_i=0.1$ on regular sides, taking $\mu_1$ as the length unit; the black-hole controls share the corresponding observer-side parameters. The $3\times4$ layout contains twelve panels. The first row compares BH--NR and NR--NR; the second compares BH--R and R--NR; the third contains R--R and the two radial intensity comparisons for the NR and R observer exteriors. In each image pair, the full view is on the left and the quadrant zoom on the right; two gray lines connect the indicated zoom region. Full and quadrant fields have sides $36\mu_1$ and $11\mu_1$, respectively. The two profile panels show $I_{\rm obs}/I_0$ over $0\le\beta_1/\mu_1\le18$, with blue solid wormhole curves, a green dash-dotted R--R curve and red dashed black-hole curves. Gray vertical markers give $\beta_{c1}$; blue or green markers give $Z\beta_{c2}$. All maps share $I_{\max}/I_0\simeq0.308$, with exponents $\gamma=0.5$ and $0.25$ for full and zoom views. The right color bar represents $D$. The first region of each wormhole is the observer and emitting side.}
\label{fig:atw-model-a-observables}
\end{figure}

\begin{figure}[!ht]
\centering
\includegraphics[width=\linewidth,height=0.79\textheight,keepaspectratio]{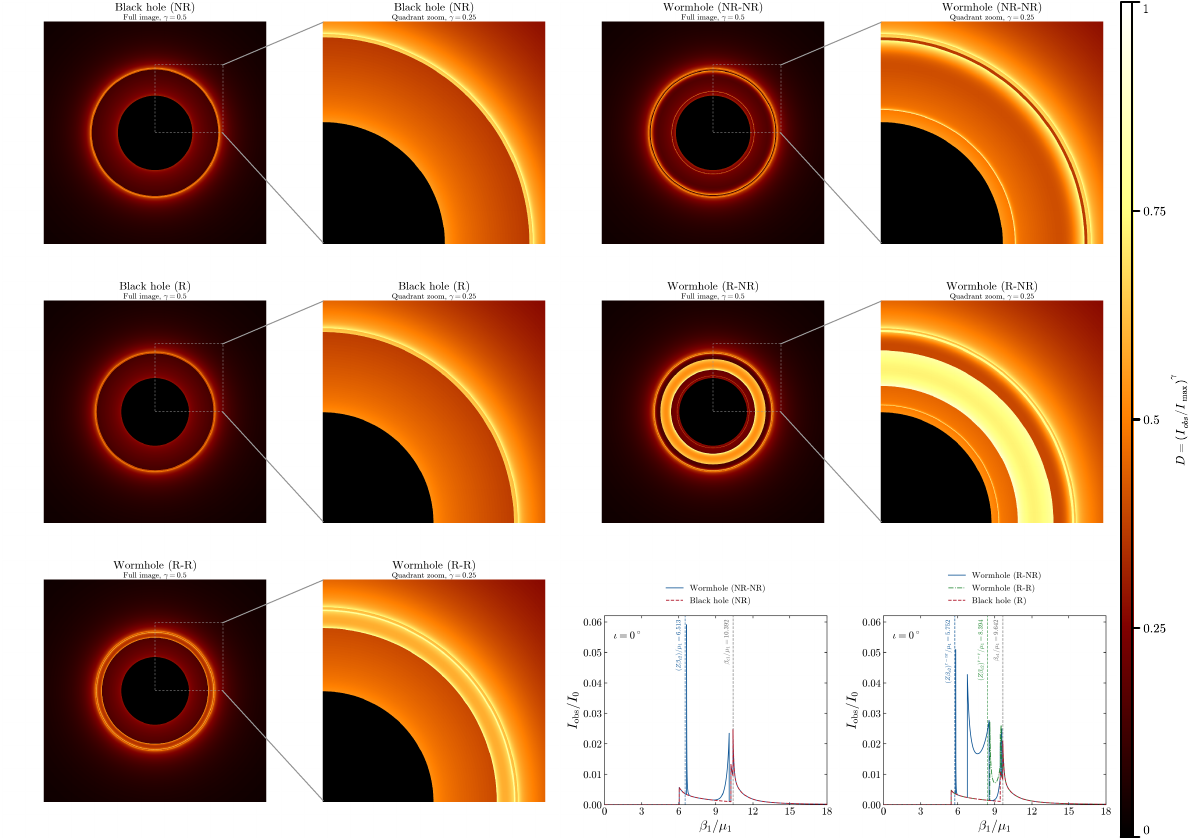}
\caption{Face-on moving-emitter comparison for Model B, whose formal inner cutoff is $r_{\rm ph}$. With $\mu_1$ as the length unit, the wormhole parameters are $\mathcal A=0.5$, $\mu_2/\mu_1=1.2$, $R/\mu_1=5.1$, and $r_{0i}/\mu_i=0.1$ on regular sides; the black-hole controls share the corresponding observer-side parameters. The $3\times4$ panel arrangement, screen fields, paired zoom connectors and critical markers follow Fig.~\ref{fig:atw-model-a-observables}. The last two panels overlay the matched NR and R radial intensities, using one ordinate range. All five configurations share $I_{\max}/I_0\simeq0.0591$. Full images and quadrant zooms use $\gamma=0.5$ and $0.25$, respectively; the right color bar shows the transformed display value $D$. The source extends into the plunging region, so the radial photon sign is retained in the frequency shift.}
\label{fig:atw-model-b-observables}
\end{figure}

\begin{figure}[!ht]
\centering
\includegraphics[width=\linewidth,height=0.79\textheight,keepaspectratio]{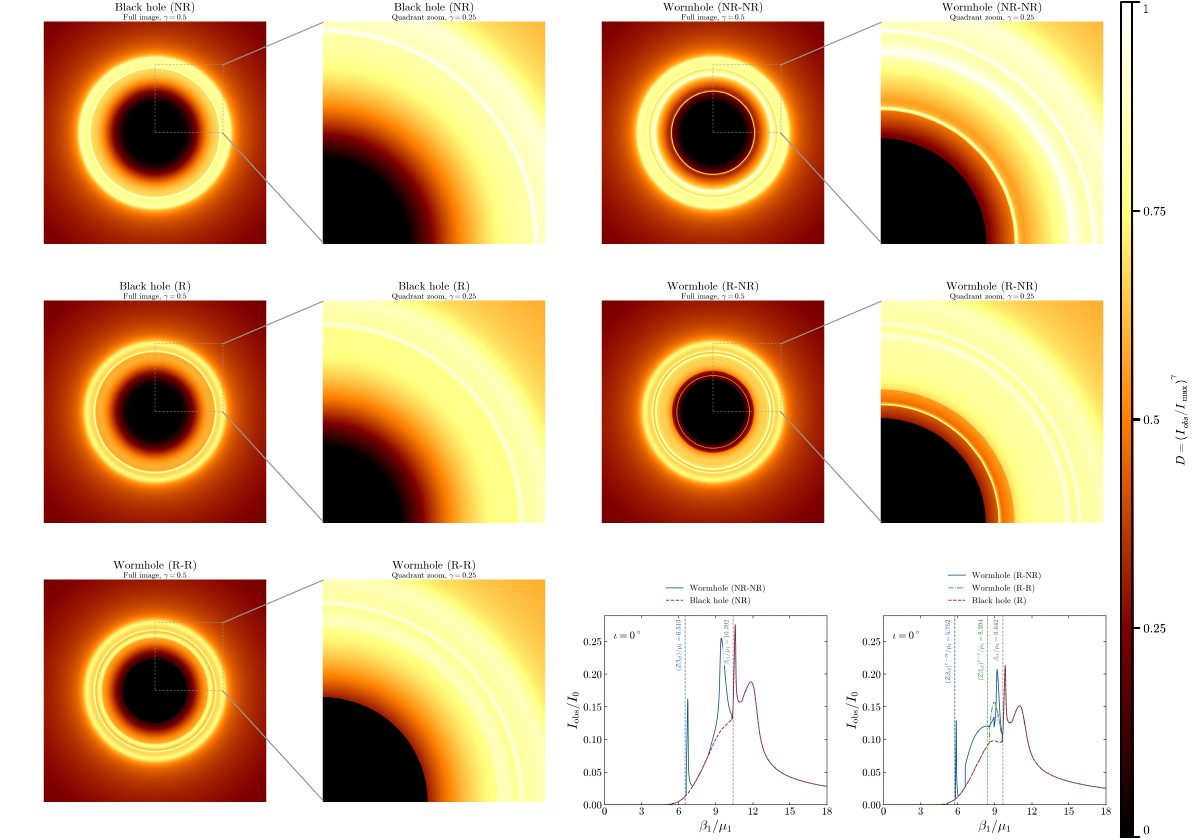}
\caption{Face-on moving-emitter comparison for Model C. With $\mu_1$ as the length unit, the wormhole parameters are $\mathcal A=0.5$, $\mu_2/\mu_1=1.2$, $R/\mu_1=5.1$, and $r_{0i}/\mu_i=0.1$ on regular sides; the black-hole controls share the corresponding observer-side parameters. The twelve-panel layout and screen fields follow Fig.~\ref{fig:atw-model-a-observables}, with common $I_{\max}/I_0\simeq0.276$. Black-hole disks retain formal support outside their horizons, whereas wormhole disks retain only $r\ge R$. The critical markers and full/zoom exponents are unchanged. All maps with the same display exponent share one physical normalization. The broad annulus and narrow rings are intensity features. The connected quadrant panels enlarge the same radial profiles; the last two panels compare the NR and R observer groups.}
\label{fig:atw-model-c-observables}
\end{figure}

Figures~\ref{fig:atw-model-a-observables}--\ref{fig:atw-model-c-observables} compare the two unexcised observer-side seed black holes with the three joined configurations, using one emission model per figure. The nonregular and regular black-hole (BH) controls, denoted BH--NR and BH--R, have the same observer-side parameters, source normalization and velocity law as the corresponding wormhole. They contain no shell or returned ray. The face-on observer has $\iota=0$, but the emitter follows Eqs.~\eqref{eq:circular-four-velocity} and \eqref{eq:plunging-four-velocity}, not a static velocity.

For each geometry, all three sources are evaluated on the same disk-crossing mesh. The face-on calculation begins with 10,400 uniformly distributed impacts over $0\le\beta_1/\mu_1\le27$, supplemented by 4,400 geometrically clustered points on each side of every applicable critical impact. Source-edge jumps and local maxima receive up to 20 successive midpoint-refinement steps. The final meshes contain 19,408, 19,657, 28,973, 28,921, 28,228 points for BH--R, BH--NR, NR--NR, R--NR and R--R, respectively. The branch quadratures use 2049 Simpson nodes per radial segment. The intensity curves and maps of a geometry are generated from the same radial profile. Maps start from $1600\times1600$ rasters: physical intensity is interpolated and transformed at twice this resolution before Lanczos downsampling. The following peak values are rounded summaries of these sampled profiles, not analytical extrema.

One maximum intensity is shared by all five geometries within each emission model. The displayed maps apply $D=(I_{\rm obs}/I_{\max})^\gamma$, where $D$ is the dimensionless display value, $I_{\max}$ is that common model maximum and $\gamma$ is a display exponent: $\gamma=0.5$ for the full image and $\gamma=0.25$ for the quadrant zoom. These stretches do not modify the intensity curves or profile integrals. Colors are quantitatively comparable among maps with the same exponent within an emission model. A given color corresponds to different physical intensities in the full and zoom views, which use different exponents.

The controls distinguish capture from absence of light. A subcritical black-hole ray can intersect emitting material before reaching the horizon; capture removes its later continuation, not those earlier contributions. Conversely, an allowed trajectory remains dark when its intersections lie outside the source support. The maps show optical intensity, not matter density, and the boundary of a dark region need not coincide with the geometric critical impact.

Model A selects only circular emitters outside the ISCO, so its face-on weights follow Eq.~\eqref{eq:polar-circular-weight}. In the nonregular control, the two isolated source-edge maxima near $\beta_1/\mu_1=10.633$ and $11.423$ reach approximately $0.250I_0$. This value agrees with the circular weight at $r_{\rm ISCO}=12\mu_1$, rather than the larger static-emitter weight. The outer total maximum near $12.578$ reaches $0.308I_0$ because image orders overlap. The corresponding regular-control features occur near $9.904$, $10.249$ and $11.718$, with the total maximum approximately $0.260I_0$. Regularization changes the ordinary transfer map, the source edge and the circular time-dilation factor together; it does not impose equality of total peaks or a universal dimming factor.

The NR--NR and R--NR configurations add inner source-edge features near $\beta_1/\mu_1=(6.771,9.294)$ and $(5.893,9.102)$, respectively. Their locations are set by returned intersections reaching the ISCO support, not by new photon spheres. R--R instead nearly coincides with BH--R throughout the displayed face-on profile: none of its retained returned targets reaches the nonzero Model A support. This is a brightness degeneracy through the first three targets, not the absence of reflected trajectories. Because the source cutoff exceeds $R$, the ordinary luminous portion of the matching black-hole control is not excised for this model. Thus the additional inner rings in the other wormholes isolate returned propagation, whereas the R--R critical marker alone does not guarantee a visible ring.

Model B samples the flow between the photon sphere and the ISCO and therefore tests the radial Doppler term that is absent from a static-emitter calculation. The largest control intensities are approximately $0.0247I_0$ for BH--NR and $0.0209I_0$ for BH--R, near their respective observer critical impacts. With the same velocity and source prescription, the wormhole maxima are approximately $0.0591I_0$ for NR--NR, $0.0510I_0$ for R--NR and $0.0271I_0$ for R--R. The brightest feature in the first two wormholes is now in the inner returned sector, near $\beta_1/\mu_1=6.631$ and $5.823$, rather than the outer observer-critical cluster.

This ordering follows from two coupled effects. The transfer map determines the emission radius and the radial segment of each returned crossing; Eq.~\eqref{eq:redshift} then assigns a different weight to that segment. For inward-moving gas, reversing the photon radial sign reverses the contribution of $-u^r k_r$ to the denominator. Returned photons can consequently be less deboosted than ordinary locally outgoing photons emitted at comparable radii. This does not mean every returned branch is bright: the source still vanishes at and below $r_{\rm ph}$, and steep transfer slopes restrict its screen support.

The two regular-observer configurations have identical local source and velocity functions, but changing $\Mtwo$ changes their crossing radii and radial segments. The R--NR configuration contains several returned structures, including features near $6.778$ and $8.536$, whereas R--R concentrates its additional peaks near $8.600$ and $9.520$. Its maximum is lower than the R--NR maximum on the common scale, despite both having brighter inner features than BH--R. This comparison isolates the second exterior more cleanly than a comparison with NR--NR, which also changes the observer-side geometry and the emitter motion. The result is specific to the adopted inward geodesic prescription and bolometric source law.

Model C reveals both the returned sector and the source excision introduced by the shell. The black-hole controls have gradually brightening inner profiles because emission persists toward the horizon, while the plunge and redshift strongly attenuate outgoing photons from the innermost radii. In the wormholes, the first observer-side target appears only when $\phi_1(R)=\pi/2$, at approximately $\beta_1/\mu_1=5.220$ for NR--NR and $5.180$ for the regular observer side. The source is nonzero at $R$, giving finite one-sided intensities of approximately $0.000552I_0$ and $0.00224I_0$, respectively, when evaluated with Eq.~\eqref{eq:redshift} on the initial radial segment. The static factor $[F_1(R)/F_\infty]^2$ is not appropriate. The much smaller moving-emitter onset follows from radial Doppler attenuation and must not be mistaken for the loss of the throat-edge branch.

The largest sampled intensities, approximately $0.276I_0$ for BH--NR and NR--NR and $0.213I_0$ for BH--R, R--NR and R--R, occur near $\beta_1/\mu_1=10.574$ and $9.838$, respectively. Equality of the largest peak within an observer-side group does not imply equality of its full image. The wormholes redistribute intensity at smaller impacts: NR--NR develops pronounced inner structure near $6.731$ and $9.456$, while R--NR and R--R differ around $5.873$ and $8.965$, among other retained features. The shared outer structure is tied to the identical observer exterior, whereas the lower-impact differences probe the shell and the opposite side.

Unlike Models A and B, the Model C black-hole source emits below $R$. Replacing the control by a wormhole therefore both removes that inner source region and introduces returned crossings. A wormhole-minus-control intensity or profile integral is consequently a net difference, not the returned component alone. This distinction is essential even when the full maps look similar: the quadrant zoom exposes changes in narrow rings superposed on the broad annulus without converting them into independent measurements.

Together, the three comparisons separate geometric accessibility, illumination and motion. Model A hides the face-on returned R--R sector; Models B and C reveal parts of it with different Doppler and radial-source weights. Peak surface brightness is not integrated flux: a narrow bright annulus occupies little screen area, and flux integration would require the additional radial measure proportional to $\beta_1\,\dd\beta_1$. No instrument-convolved or area-weighted flux is inferred from these maps. The results demonstrate a controlled response to the regularization assignment within this thin-disk model, not a source- or velocity-independent identification of a wormhole.

\subsubsection{Inclination dependence}
\label{sec:inclined-disk}

The face-on results in Sec.~\ref{subsec:joined-face-on} establish how source support selects the visible returned branches, including the near-degeneracy of the R--R Model A image with its black-hole control. We now examine which brightness conclusions survive a change of viewing angle. We compare $\iota=17^\circ,45^\circ,80^\circ$ with the same spacetime parameters and the circular-plus-plunging motion prescribed in Sec.~\ref{subsec:thin-disk-emission}. Inclination changes both the disk intersections and their frequency weights. The approaching side, on the left in our screen convention, becomes increasingly brighter than the receding side; the circular symmetry of the face-on brightness distribution is lost.

\begin{figure}[H]
\centering
\includegraphics[width=\linewidth]{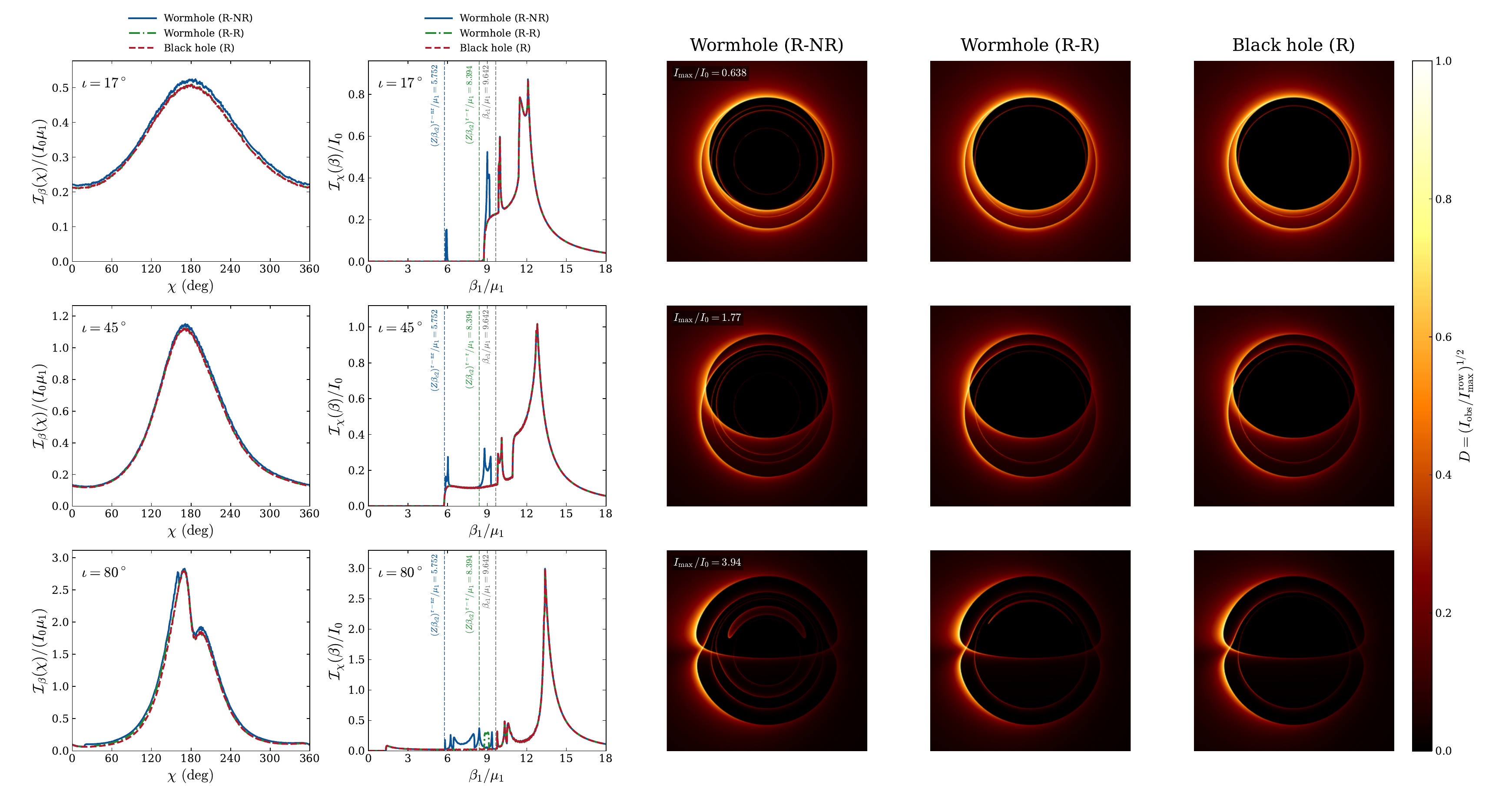}
\caption{Model A at $\iota=17^\circ,45^\circ,80^\circ$ (rows), comparing R--NR, R--R, and the regular black hole for wormhole parameters $\mathcal A=0.5$, $\mu_2/\mu_1=1.2$, $R/\mu_1=5.1$, and $r_{0i}/\mu_i=0.1$ on regular sides, taking $\mu_1$ as the length unit; the black-hole control shares the observer-side parameters. The first two columns show $\mathcal I_\beta(\chi)/(I_0\mu_1)$ and $\mathcal I_\chi(\beta)/I_0$; the last three show the corresponding images in a square screen field of side $36\mu_1$. Vertical dashed lines indicate the critical scales in Table~\ref{tab:critical}. All maps in a row share $D=(I_{\rm obs}/I_{\max}^{\rm row})^{1/2}$. The additional R--R structure at $80^\circ$ lifts the lower-inclination near-degeneracy without shifting the dominant peak.}
\label{fig:inclination-regular-a}
\end{figure}

For a fixed emission model, write $I_{\rm obs}(\beta,\chi;\iota)\equiv\mathscr I_\iota^{(k)}(\beta_1,\chi)$, with $\beta=\beta_1$. Two complementary profiles characterize the noncircular image:
\begin{align}
\mathcal I_\chi(\beta;\iota)
&=\int_0^{2\pi}I_{\rm obs}(\beta,\chi;\iota)\,d\chi,
\label{eq:inclination-angular-integral}\\
\mathcal I_\beta(\chi;\iota)
&=\int_0^{18\mu_1}I_{\rm obs}(\beta,\chi;\iota)\,d\beta.
\label{eq:inclination-radial-integral}
\end{align}
The first resolves radial structure after angular integration; the second retains the azimuthal distribution after radial integration. We plot $\mathcal I_\chi/I_0$ and $\mathcal I_\beta/(I_0\mu_1)$. These profile integrals have no radial area weight and are not total observed fluxes. Images in each row share the display stretch $D=(I_{\rm obs}/I_{\max}^{\rm row})^{1/2}$, so colors can be compared within a row but not between inclinations. Three representative comparisons are retained below; the complementary geometry--model combinations and numerical rendering details are given in Appendix~\ref{app:inclination-comparisons}.

\begin{figure}[H]
\centering
\includegraphics[width=\linewidth]{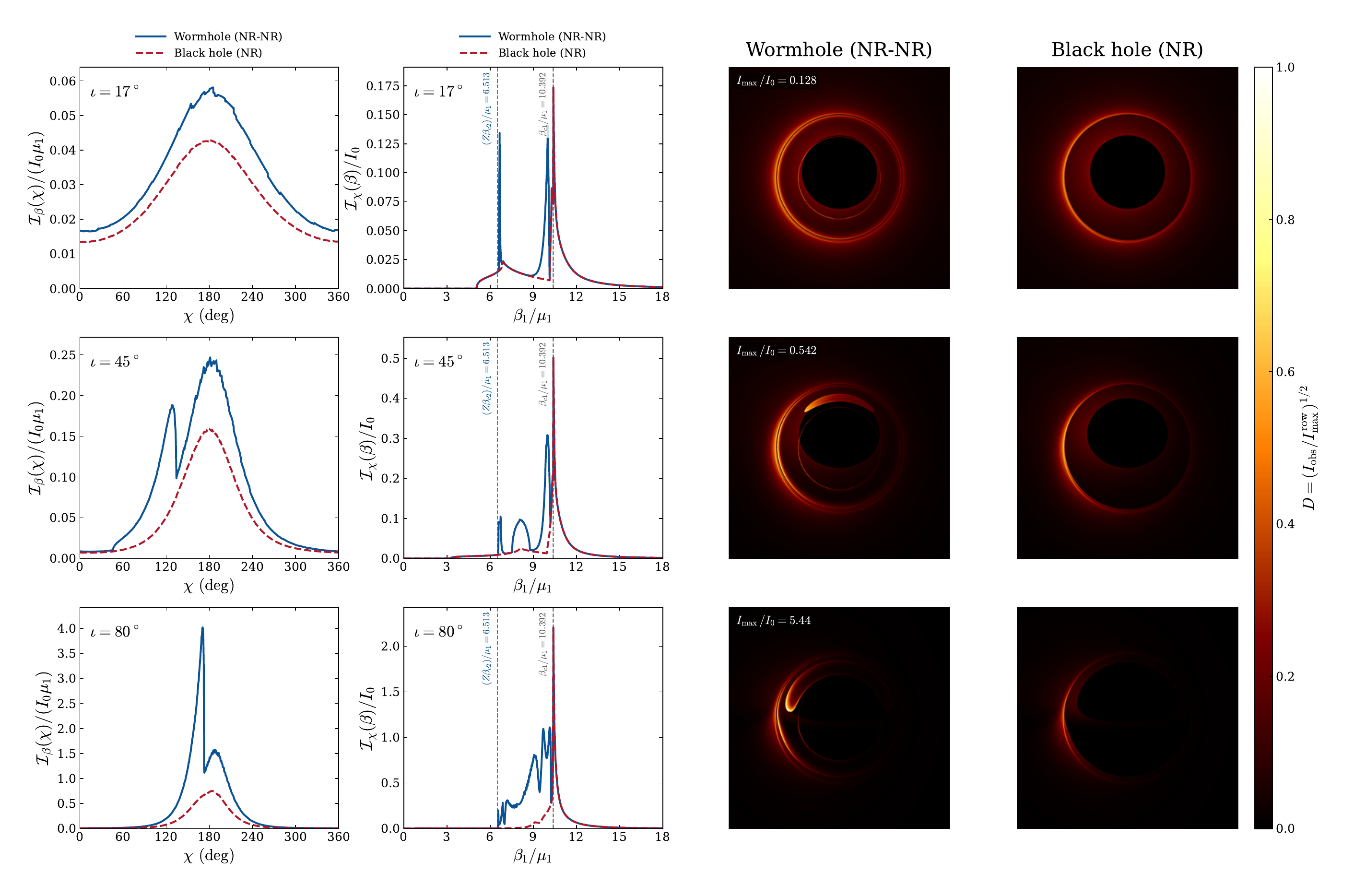}
\caption{Model B inclination comparison between NR--NR and the nonregular black hole. With $\mu_1$ as the length unit, the NR--NR wormhole uses $\mathcal A=0.5$, $\mu_2/\mu_1=1.2$, and $R/\mu_1=5.1$; the nonregular black-hole control shares its observer-side parameters. Rows use $\iota=17^\circ,45^\circ,80^\circ$. Rows, profile definitions, image domains, and within-row normalization follow Fig.~\ref{fig:inclination-regular-a}; the last two columns contain the wormhole and black-hole maps. The angular profile at $45^\circ$ reveals a wormhole secondary peak near $\chi\simeq128^\circ$ despite nearly equal integrated left--right ratios. The dashed radial markers are the corresponding NR critical scales in Table~\ref{tab:critical}.}
\label{fig:inclination-nr-b}
\end{figure}

The critical scales and ray classes derived above remain unchanged. What changes relative to the face-on view is the angular dependence of the illumination: a branch may have a peak radius $\beta_{\rm p}(\chi)$, and angular integration superposes these differently placed and Doppler-weighted contributions. A narrow face-on feature can therefore broaden or split within the existing geometric support, without uniformly filling the interval between the critical radii in Table~\ref{tab:critical}. A branch that becomes newly illuminated must instead be distinguished from the broadening of an already bright feature.

The most direct change to the face-on conclusions occurs in Model A, shown in Fig.~\ref{fig:inclination-regular-a}. R--R remains nearly coincident with the regular black-hole control at $17^\circ$ and $45^\circ$, with no returned emission from the retained crossings. At $80^\circ$, a returned feature appears near $\beta_1/\mu_1\simeq9.05$, while the dominant radial peak remains coincident with the black-hole peak. The additional emission supplies approximately $2.1\%$ of $\int\mathcal I_\chi\,\dd\beta_1$ over the displayed window.

The illumination change follows directly from Eq.~\eqref{eq:first-disk-crossing}. For $\sin\chi>0$, increasing $\iota$ moves the targets $\phi_m$ to later accumulated angles. On the observer-side return segment, the intersection satisfies $\phi_{1,{\rm in}}(r_m;\beta_1)=\phi_{\rm tot}-\phi_m$. Since the incoming angle decreases with radius, a later target moves this returned intersection outward. In the sampled R--R polar grid, the largest returned radii among the first three global targets are approximately $6.623\mu_1$, $7.149\mu_1$ and $8.452\mu_1$ at $\iota=0^\circ,17^\circ,45^\circ$, respectively. All lie below the Model A cutoff $r_{\rm ISCO}=10.715\mu_1$, so their emitted intensity vanishes. At $80^\circ$, third-target returned intersections reach $11.827\mu_1$ and enter the emitting region. The reflected impact interval in Table~\ref{tab:critical} is unchanged; inclination changes which part of its return segment intersects the source.

The source-support condition favors $\sin\chi>0$, whereas the approaching Doppler enhancement favors $\cos\chi<0$, for which $k_\phi>0$ in Eq.~\eqref{eq:photon-momentum-at-disk}. Their overlap concentrates the extra emission in the upper approaching sector. The Doppler factor weights crossings with nonzero $I_{\rm em}^{(A)}$; it cannot illuminate a crossing below the ISCO cutoff. Since $r_{\rm ISCO}>R$, the matched geometries share their ordinary luminous source and circular velocity law. The black-hole path ends at capture and has no corresponding return segment, while wormhole photons can visit $\Mtwo$ and return to emitting matter in $\Mone$. This geometric access to the source, followed by the circular Doppler weighting, lifts the face-on brightness degeneracy through the retained third target. The sampled inclinations do not establish a universal onset angle, and the reported profile integral has no radial area weight.

The azimuthal profile makes this distinction more specific. The R--R excess is largest near $\chi\simeq142^\circ$, whereas the dominant angular peak remains near $170^\circ$ in both geometries. A horizontal cut would miss the additional structure. Moreover, the ratio of the angular integrals of $\mathcal I_\beta$ over the left half-screen ($90^\circ<\chi<270^\circ$) and the complementary right half-screen decreases slightly, from $10.25$ for the black hole to $9.89$ for R--R. Additional returned emission therefore does not necessarily strengthen the overall left--right contrast.

\begin{figure}[H]
\centering
\includegraphics[width=\linewidth]{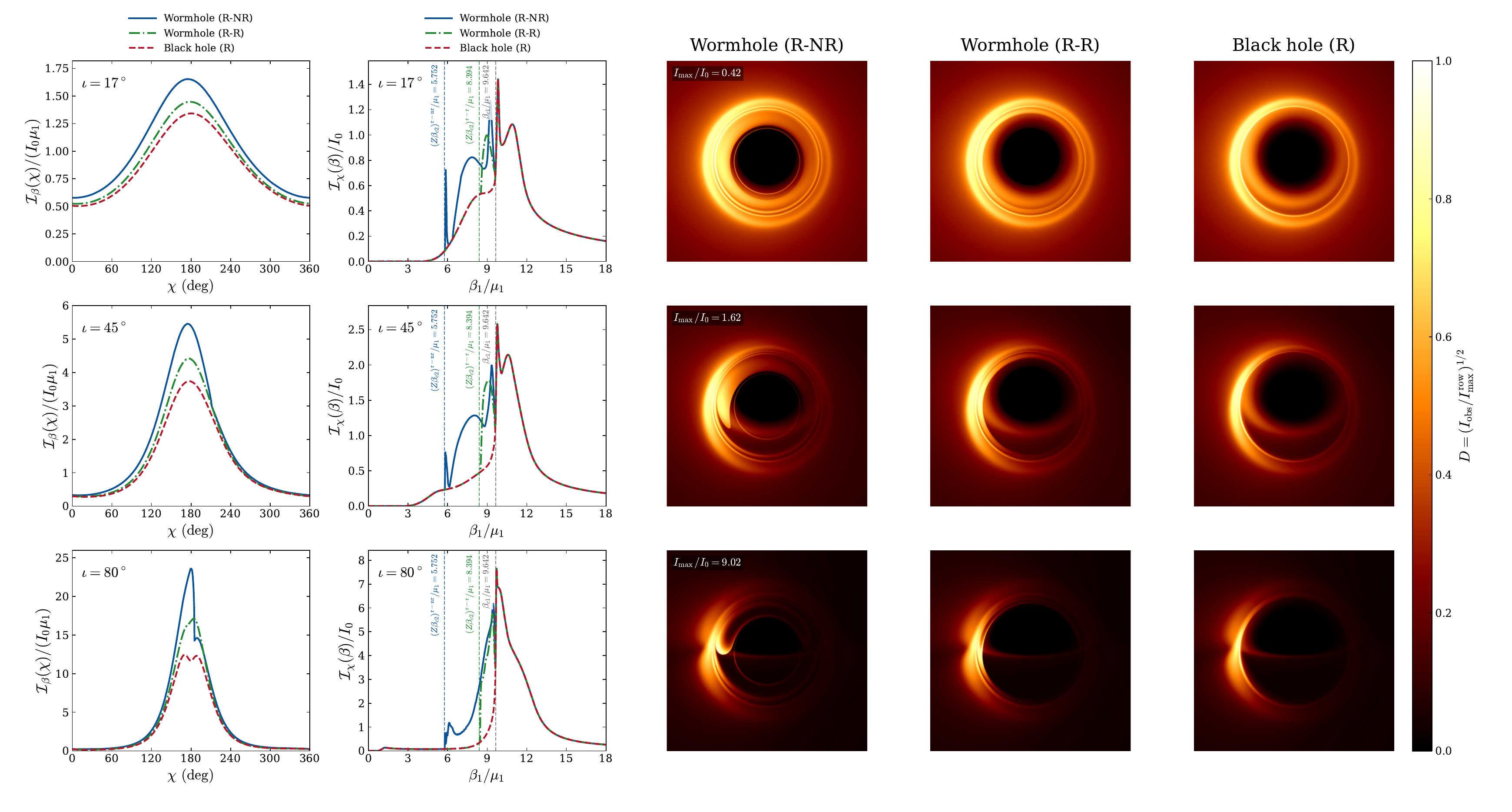}
\caption{Model C inclination comparison for R--NR, R--R, and the regular black hole, with the same layout and normalization as Fig.~\ref{fig:inclination-regular-a}. With $\mu_1$ as the length unit, the wormhole parameters are $\mathcal A=0.5$, $\mu_2/\mu_1=1.2$, $R/\mu_1=5.1$, and $r_{0i}/\mu_i=0.1$ on regular sides; the black-hole controls share the corresponding observer-side parameters. Rows use $\iota=17^\circ,45^\circ,80^\circ$. At $80^\circ$, the different returned contributions reweight the two bright-side maxima of the black-hole angular profile, yielding different dominant-peak locations in the two wormholes. The radial and angular profiles are computed before the display stretch.}
\label{fig:inclination-regular-c}
\end{figure}

Model B illustrates a different effect of inclination (Fig.~\ref{fig:inclination-nr-b}). Its returned emission, already visible face-on, becomes more strongly weighted at large inclination: the NR--NR excess in $\int\mathcal I_\chi\,d\beta$ over the nonregular black hole increases from approximately $29\%$ at $17^\circ$ to $181\%$ at $80^\circ$. Both integrals increase, so this relative growth reflects a faster increase of the returned contribution rather than a dimming control. This trend is not universal: the corresponding Model A excess is only $5.7\%,6.0\%,3.3\%$ across the three inclinations (Appendix~\ref{app:inclination-comparisons}). Stronger Doppler asymmetry alone is therefore insufficient to predict a stronger wormhole--black-hole difference.

Inclination also provides an angular discriminator absent from a circular face-on image. At $45^\circ$, the NR--NR Model B profile develops a secondary peak near $\chi\simeq128^\circ$ that is absent from the black-hole control, although their dominant peaks remain near $180^\circ$ and their left-to-right ratios are almost equal ($7.72$ and $7.63$). Returned propagation can redistribute brightness within the same bright half-screen without strongly changing that integrated ratio. The shape of $\mathcal I_\beta(\chi)$ consequently carries information missed by either a radial profile or a single asymmetry measure.

Model C shows how this angular information separates the two regular-observer wormholes (Fig.~\ref{fig:inclination-regular-c}). At $80^\circ$, the regular black-hole profile has two local maxima of similar height near $170^\circ$ and $190^\circ$. R--NR preferentially enhances the region between them, placing the dominant maximum near $180^\circ$, whereas R--R peaks near $185^\circ$. Unlike the concentric inner-ring differences in the face-on maps, this distinction concerns the relative weights of features within the approaching half-screen. Since the observer-side source and velocity field are shared, it follows from the different returned crossing maps and frequency weights, rather than a change in disk rotation. As in the face-on comparison, the Model C wormhole--black-hole difference also includes the truncation of the source at the throat; it is not exactly the returned intensity.

The new inclination-dependent conclusions are therefore the lifting of a brightness degeneracy, the possible broadening of radial features, and the redistribution of angular peaks and shoulders. These effects complement the fixed geometric classification established earlier. Their strength remains dependent on the source prescription, and the present idealized images do not establish observational distinguishability.

\raggedbottom

\section{Conclusion}\label{sec:conclusion}

We have investigated how a rational regularization of the new Letelier--Alencar string-cloud geometry changes the optics of a reflection-asymmetric thin-shell wormhole. The comparison was organized among nonregular--nonregular, regular--nonregular, and regular--regular exterior assignments while holding the string-cloud strength, throat radius, regulator fractions, and leading asymptotic masses fixed. Matching the asymptotic masses is central to the interpretation: it removes the leading inverse-radius mass shift and attributes the remaining differences to the finite-radius structure of the regularized lapse and to its role in the shell matching.

Although the regular anti--de Sitter core is excised by the throat and is never sampled directly by the calculated rays, regularization has observable consequences within the idealized optical model because it modifies the retained exterior. For the sampled parameters, regularizing the observer side moves the horizon, photon sphere, innermost stable circular orbit, and critical screen impact inward; the latter decreases from $10.392\mu_1$ to $9.642\mu_1$. Regularizing the second side leaves the ordinary observer-side critical curve unchanged but raises its mapped critical edge from $5.752\mu_1$ to $8.394\mu_1$, contracting the reflected screen interval from $3.890\mu_1$ to $1.248\mu_1$. The comparison therefore distinguishes two mechanisms: observer-side regularization changes both ordinary propagation and shell matching, whereas second-side regularization acts specifically on crossing rays through the matching factor and the second-side angular integral.

This separation is visible in the ray classes, orbit numbers, and transfer functions. At fixed observer-screen impacts, changing the regularization assignment converts transmitted rays into reflected rays, reflected rays into transmitted rays, and reflected rays into ordinary scattered rays without retuning the incident light. The returned orbit-number families respond to both sides, while trajectories confined to the observer exterior are insensitive to changes made only across the throat. For the largest sampled string-cloud strength, regularizing the second exterior substantially narrows the returned third-order support and removes the retained reflected second-order disk intersection. 
These statements describe the sampled parameter sets and are not global classifications of every regulator strength or throat position.

The moving-emitter comparisons show that geometric accessibility and brightness cannot be identified with one another. Matched black-hole and wormhole images share their source and velocity prescriptions, including the radial Doppler term below the innermost stable circular orbit. Model A hides the retained face-on returned sector in the regular--regular case, whereas Models B and C reveal additional inner structures with different frequency weights. Model C also probes the removal of emitting radii below the throat, so its wormhole-minus-control difference is not a pure returned intensity. Peak surface brightness is distinct from area-integrated flux. Inclination changes disk intersections and Doppler weights without moving the geometric critical scales: at the largest sampled angle, the Model A regular--regular near-degeneracy with its black-hole control is lifted, and the other models show redistributed radial and angular structure. These results support a regularization-sensitive optical response under the stated assumptions, not a unique, source-independent or presently detectable signature.

The analysis assumes static spherical shells, an optically thin test disk with isotropic bolometric emission, a specified circular-plus-ISCO-plunge velocity law, and only the first three disk-plane targets. The optical families use one fixed set of geometric parameters and a three-point string-strength scan, while inclination is sampled at three angles; neither higher-order intensity nor a universal inclination threshold is established. Pressure, magnetic stresses, absorption, plasma propagation and time-dependent accretion are not modeled. An important extension is to repeat the matched comparison with alternative regularization prescriptions, enforcing the same asymptotic mass, throat conditions, source and screen normalization to distinguish regulator-specific effects from generic consequences of nonsingularity. Broader scans of regulator scale, string strength, mass ratio and throat radius should also be combined with shell-stability constraints and higher image orders. Alternative velocity fields, dynamical or rotating shells and instrument-level image synthesis are needed before these idealized differences can be assessed as observational discriminants.

\begin{acknowledgments}
Edson Otoniel would like to thank Fundação Cearense de Apoio ao Desenvolvimento Científico e Tecnológico through grant BP6-0241-00335.01.00/25. This work was supported by NSFC under Grant No.~12575048. M.G. also acknowledges support from the BNU Tang
Scholar Program.
\end{acknowledgments}

\appendix
\section{Complementary inclination comparisons}
\label{app:inclination-comparisons}

\begin{figure}[H]
\centering
\includegraphics[width=\linewidth]{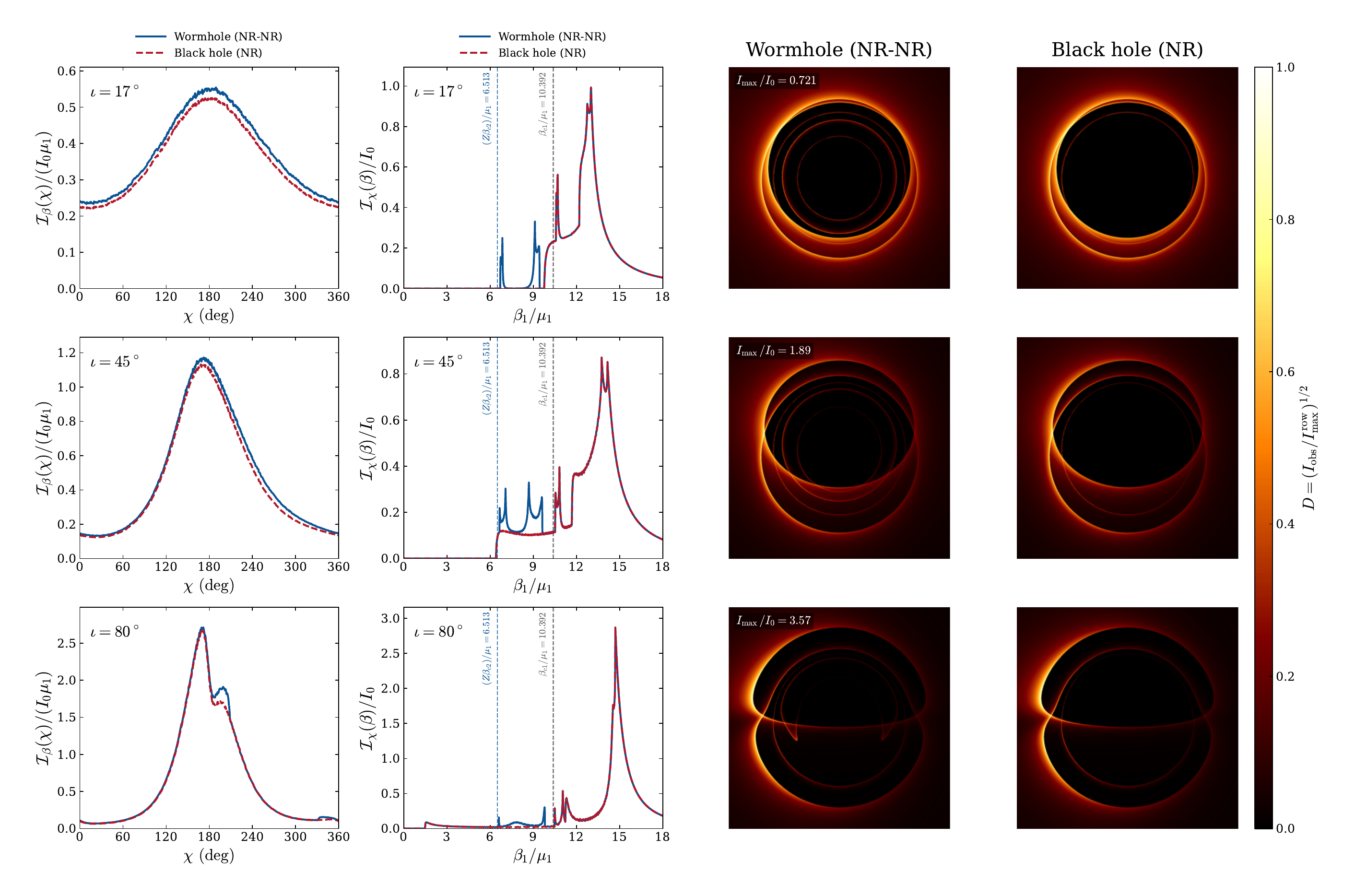}
\caption{Complementary Model A comparison between NR--NR and the nonregular black hole. With $\mu_1$ as the length unit, the NR--NR wormhole uses $\mathcal A=0.5$, $\mu_2/\mu_1=1.2$, and $R/\mu_1=5.1$; the nonregular black-hole control shares its observer-side parameters. Rows use $\iota=17^\circ,45^\circ,80^\circ$. Rows, profile definitions, map domains, critical markers, and within-row normalization follow the NR layout of Fig.~\ref{fig:inclination-nr-b}. The returned contribution is present at all three inclinations, while its relative integral weight is smallest at $80^\circ$.}
\label{fig:inclination-nr-a}
\end{figure}

\begin{figure}[H]
\centering
\includegraphics[width=0.9\linewidth]{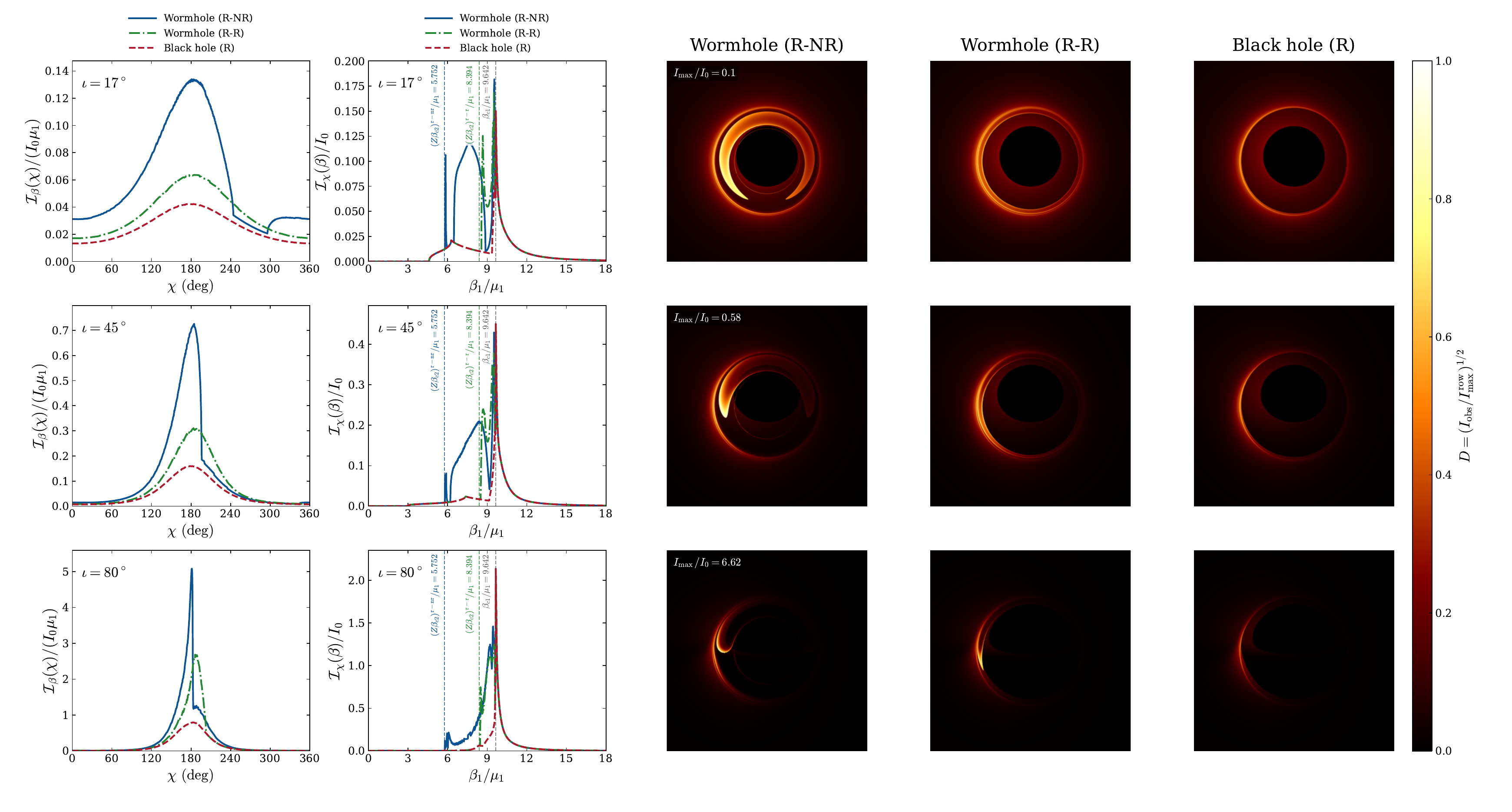}
\caption{Complementary Model B comparison of R--NR, R--R, and the regular black hole. With $\mu_1$ as the length unit, the wormhole parameters are $\mathcal A=0.5$, $\mu_2/\mu_1=1.2$, $R/\mu_1=5.1$, and $r_{0i}/\mu_i=0.1$ on regular sides; the black-hole controls share the corresponding observer-side parameters. Rows use $\iota=17^\circ,45^\circ,80^\circ$. Layout, domains, and normalization follow Fig.~\ref{fig:inclination-regular-a}. Both wormholes already have additional returned emission at the lowest inclination; increasing inclination changes its weight and angular concentration.}
\label{fig:inclination-regular-b}
\end{figure}

\begin{figure}[H]
\centering
\includegraphics[width=0.8\linewidth]{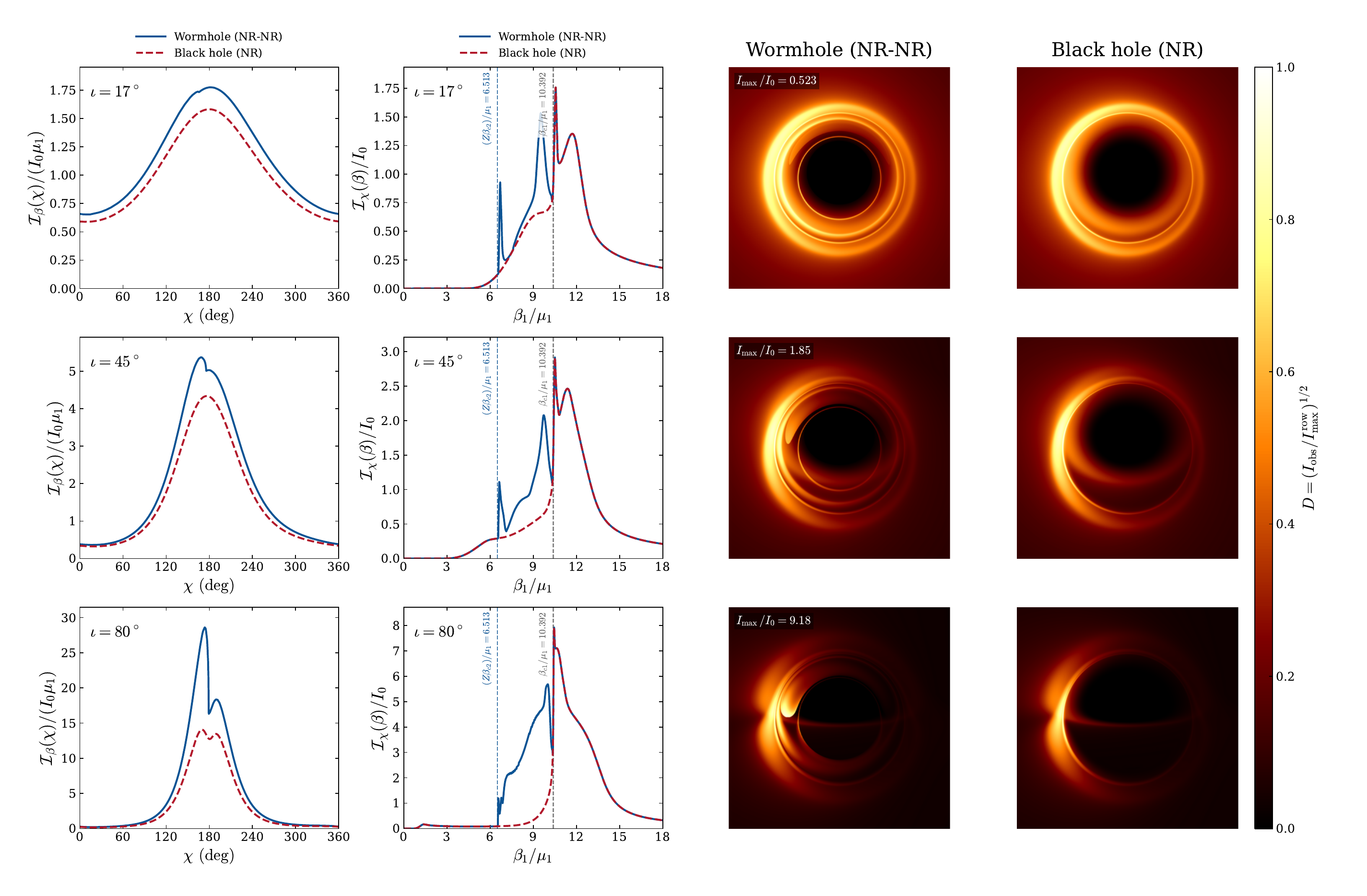}
\caption{Complementary Model C comparison between NR--NR and the nonregular black hole, with the layout and conventions of Fig.~\ref{fig:inclination-nr-b}. With $\mu_1$ as the length unit, the NR--NR wormhole uses $\mathcal A=0.5$, $\mu_2/\mu_1=1.2$, and $R/\mu_1=5.1$; the nonregular black-hole control shares its observer-side parameters. Rows use $\iota=17^\circ,45^\circ,80^\circ$. The high-inclination angular profile illustrates the unequal enhancement of the two bright-side maxima.}
\label{fig:inclination-nr-c}
\end{figure}

Figures~\ref{fig:inclination-nr-a}, \ref{fig:inclination-regular-b} and \ref{fig:inclination-nr-c} complete the three emission models and two matched observer-side geometry groups in Sec.~\ref{sec:inclined-disk}. They use the emitter prescription in Sec.~\ref{subsec:thin-disk-emission}, $\iota=17^\circ,45^\circ,80^\circ$, and the first three disk-plane targets. The native polar sampling extends to $\beta_1/\mu_1=27$ to cover the corners of the square maps with half-width $18\mu_1$; profile integrals use only $0\le\beta_1/\mu_1\le18$. Display interpolation and antialiasing do not modify those integrals.

For Model A, the NR--NR excess in the unweighted profile integral is $5.7\%,6.0\%,3.3\%$ at the three inclinations. Thus its already visible returned component does not become monotonically more prominent relative to the black-hole control as the disk is inclined, in contrast to the high-inclination appearance of the R--R feature in the main text.

For Model B, returned emission accounts for approximately $29\%$, $40\%$, and $47\%$ of the R--R profile integral at inclinations of $17^\circ$, $45^\circ$, and $80^\circ$, respectively. These fractions are obtained by subtracting the matched BH--R integral and normalizing the difference by the total R--R integral over the same screen window. Because this source has no emission below $R$, the difference isolates the retained returned component. At $80^\circ$, the left-to-right profile-integral ratios are approximately $61.3$, $64.3$, and $36.7$ for R--NR, R--R, and the regular black hole. R--R therefore has a slightly larger asymmetry than R--NR despite a smaller total returned contribution. The returning intensity and its angular concentration need not rank the geometries in the same order.

For Model C at $80^\circ$, NR--NR preferentially enhances the smaller-$\chi$ bright-side maximum, with its dominant angular peak near $174^\circ$. This complements the regular-observer comparison in Fig.~\ref{fig:inclination-regular-c}, where the two wormholes instead peak near $180^\circ$ and $185^\circ$. These approximate positions characterize the sampled intensity profiles, not geometric critical angles; comparing NR and R observer exteriors also changes the emitting radii and velocity field.

\FloatBarrier
\bibliography{references}

\end{document}